\documentclass[bibyear]{aa}

\usepackage{float}
\usepackage{subcaption}
\usepackage{graphicx}
\usepackage[svgnames]{xcolor}
\usepackage{natbib}
\usepackage{amsmath}
\usepackage[varg]{txfonts}
\usepackage{caption}

\bibpunct{(}{)}{;}{a}{}{,}

\begin{document}

  \title{Impact of galactic shear and stellar feedback on star formation}

  \author{C\'edric Colling\inst{\ref{inst1}} 
    \and Patrick Hennebelle\inst{\ref{inst1},\ref{inst2}}
    \and Sam Geen \inst{\ref{inst3}}
    \and Olivier Iffrig \inst{\ref{inst1},\ref{inst4}}
    \and Fr\'ed\'eric Bournaud\inst{\ref{inst1}}}

  \institute{Laboratoire AIM, Paris-Saclay, CEA/IRFU/DAp -- CNRS --
    Universit\'e Paris Diderot, 91191 Gif-sur-Yvette Cedex, France
    \label{inst1}
    \and
    LERMA (UMR CNRS 8112), Ecole Normale Sup\'erieure, 75231 Paris Cedex, France
    \label{inst2}
    \and 
    Maison de la Simulation, CEA, CNRS, Univ. Paris-Sud, UVSQ, Universit\'e Paris-Saclay, F-91191 Gif-sur-Yvette
    \label{inst4}
    \and
    Zentrum Fur Astronomie der Universitat Heidelberg, Institut fur Theoretische Astrophysik, Albert-Ueberle-Strasse 2, 69120 Heidelberg, Germany
    \label{inst3}
    }

  \date{Received ---; accepted ---}

  \abstract
    {Feedback processes and the galactic shear regulate star formation.}
    {We investigate the effects of differential galactic rotation and stellar feedback on the interstellar medium (ISM) and on the star formation rate (SFR).}
    {%
      A numerical shearing box is used to perform three-dimensional
      simulations of a $1\ \mathrm{kpc}$ stratified cubic box of turbulent and
      self-gravitating interstellar medium (in a rotating frame) with supernovae and HII feedback. We vary the value of
      the velocity gradient induced by the shear and the initial value of the galactic magnetic field. 
      Finally the different star formation rates 
      and the properties of the structures associated with this set of simulations are computed.
    }
    {We first confirm that the feedback has a strong limiting effect on star formation.
    The galactic shear has also a great influence: the higher the shear, the lower the SFR. 
    Taking the value of the velocity gradient in the solar neighbourhood, the SFR is too high compared to the observed Kennicutt law, by a factor approximately three to six.
    This discrepancy can be solved by arguing that the relevant value of the shear is not the one in the solar neighbourhood, 
    and that in reality the star formation efficiency within clusters is not $100 \%$.
    Taking into account the fact that star-forming clouds generally lie in spiral arms where the shear can be substantially higher (as probed by galaxy-scale simulations), 
    the SFR is now close to the observed one.
    Different numerical recipes have been tested for the sink particles, giving a numerical incertitude of a factor of about 
    two on the SFR.
    Finally we have also estimated  the velocity dispersions in our dense clouds and found that they lie below the observed Larson law
    by a factor of about two.}
    {In our simulations, magnetic field, shear, HII regions, and supernovae all contribute significantly to reduce the SFR.
     In this numerical setup with feedback from supernovae and HII regions and a relevant value of galactic shear, 
     the SFRs are compatible with those observed, with a numerical incertitude factor of about two.} 
  \keywords{ISM: clouds - ISM: magnetic fields - ISM: structure - Stars: formation}
  \maketitle
\section{Introduction}

Star formation, a key phenomenon in our universe, is still not completely understood. 
Part of the reason for this is that many physical processes interact over a large range of temporal and spatial scales.
Our understanding of star formation is directly linked to our comprehension of the cycle and the dynamics of the interstellar medium (ISM).
Because of this large range of scales, it is not possible to simulate a galaxy 
\citep[e.g.][]{tasker-bryan2006,dubois+2008,bournaud+2010,kim+2011,dobbs+2011,tasker2011,hopkins+2011,Bournaud2013,Goldbaum2016,Semenov2017} 
with a well-resolved interstellar medium everywhere. A resolution of $\sim 2 \ \mathrm{pc}$ is possible with adaptive mesh refinement, but only in very dense regions. 
One possible solution is to simulate a small part of a galaxy in order to have a better spatial resolution 
\citep[e.g.][]{korpi+1999,slyz+2005,deavillez+2005,Joung06,Hill12, kim+2011, kim+2013, gent+2013, Hennebelle14,gatto2015} 
at the cost of not solving the large galactic scales.

In this paper, we study the influence of stellar feedback (by supernovae and HII regions) and the galactic shear
on the ISM and the star formation rate (SFR) by performing a series of simulations 
of a stratified \textbf{shearing box}, with various physical parameters such as the initial magnetic field or the velocity gradient in the box.
The goal is to reproduce the observed Kennicutt law for the SFR \citep{Kennicutt12}.
It is already known that magnetic fields and turbulence have a limiting effect on star formation \citep{Ostriker2011, Hennebelle12}.
Feedback from stars can also greatly reduce the SFR \citep{Hennebelle14, Kim15b, Gatto2017}.
The values of the SFR obtained in \citet{Iffrig2017} (with only supernovae feedback) were actually close to the observed ones.
Supernovae inject momentum and energy into the gas around the dying star.
In this way, they are capable of unbinding the gas in the host cloud of the massive star.
The problem was that in these models, the supernovae directly occurred at the creation of the stars, and thus were not fully realistic. \\
Our new model for feedback introduces a delay for the supernovae, and HII regions.
A HII region is composed of hydrogen that has been ionised by UV photons, and has a
temperature of about $10^4 \ \mathrm{K}$. Because of the temperature difference with
the surrounding neutral gas, a pressure difference is created across the
ionisation front that triggers the expansion of the HII region.
This expanding wave perturbs the interstellar medium, and therefore can prevent star formation.
Theoretical features of HII regions have been previously described in many papers 
such as \citet{Kahn1954}, \citet{Franco1990}, and \citet{Matzner2002}. 

Galactic shear is another process that competes with gravity and slows down star formation.
Previous work with simulations of shearing galactic gas disks has been done \citep[e.g.][]{Ostriker2002,Kim13}.
\citet{Kim13} find that with a gas column density $\Sigma_{gas} \propto \Omega$ (where $\Omega$ is the galactic angular velocity)
so that the Toomre criterion $Q$ is constant at $\sim 2$, the SFR follows the law $\Sigma_{SFR} \propto \Sigma_{gas}^2$.
\citet{Kim17} find a numerically converged SFR $\Sigma_{SFR} \sim 5 \cdot 10^{-3} \ \mathrm{M_{\odot} \, kpc^{-2} \, yr^{-1}}$
for a gas surface density $\Sigma_{gas} \sim 10 \ \mathrm{M_{\odot} \, pc^{-2}}$.

Section 2 describes the numerical and physical setup that we use, as well as the
shearing box we have implemented. In the third section, we give the results we obtained from the
various simulations, in terms of SFR and structures properties. The fourth section is a discussion about
the relevant shear value given the surface density, and about the physical limits of the model.
The fifth section concludes this paper.
\section{Numerical methods}

\subsection{Numerical code}

Our simulations are performed with a modified version of the RAMSES code \citep{Teyssier02},
a code using the Godunov method to solve the magnetohydrodynamics (MHD) equations \citep{Fromang06}. We use a $256^3$ and $512^3$ grid with no adaptive mesh refinement. 
For a $1 \ \mathrm{kpc^3}$ box, that corresponds to a physical resolution
of $\sim 4$ or $\sim 2$ $\mathrm{pc}$.
The development of the shearing box also lead us to use a planar decomposition (along the $z$-axis) of the cubic domain,
each partial domain being attributed to one core.
We use sink particles \citep{Bleuler2014} to follow the evolution of the dense gas and model star formation.
\subsection{Equations}
\label{sec:eq}
  \begin{figure}
    \begin{center}
      \includegraphics[width=0.45\textwidth]{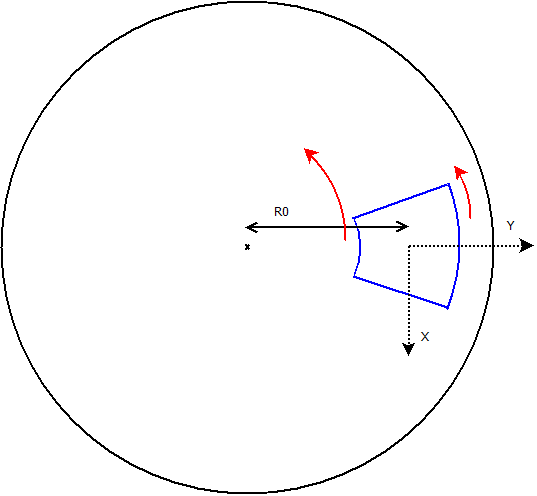}
    \end{center}
    \caption{Sketch showing the galactic differential rotation and our rotating frame.}
    \label{fig:shear}
  \end{figure}
We solve the ideal MHD equations with self-gravity, 
heating, and cooling processes, in a rotating frame (at $\Omega_0 = \Omega(R_0)$),
where $R_0$ is the distance from the centre of the box to the centre of the galaxy. 
In the galaxy, there is differential rotation, that is, $\Omega = \Omega(R)$ (see Fig. \ref{fig:shear}).
In the rotating frame, this implies for the velocity along the $x$-axis (the minus sign comes from the fact that here the $x$-axis
is in the opposite direction to the usual polar axis)
\begin{equation}
V(y)  = -\left(R_0+y\right) \cdot \left(\Omega\left(R_0+y\right)-\Omega_0\right) = -\Omega'(R_0)yR_0,
\end{equation}
after a Taylor expansion (the box is assumed to be far enough from the galactic centre). 
In this approximation, there is a constant gradient of velocity along $y$.
We define the velocity gradient 
\begin{equation}
V_{shear}~\equiv~-\Omega'(R_0)R_0=\frac{1}{L}\left(V\left(+\frac{L}{2}\right)~-~V\left(-\frac{L}{2}\right)\right)~=~\frac{2}{L} \cdot V\left(+\frac{L}{2}\right), 
\end{equation}
with $L$ being the size of the box. 

Following \citep{Ostriker2002}, assuming the angular velocity is a power law $\Omega(R) \propto R^{-q}$, 
the effective tidal potential is 
  \begin{align}
\boldsymbol{\nabla}\phi_{eff}(R = R_0+y) & = \boldsymbol{\nabla}\phi_{rot} + \boldsymbol{f}_{centrifugal}, \nonumber \\
                            & = -R\Omega^2\boldsymbol{e_y} + R\Omega_{0}^2\boldsymbol{e_y}, \nonumber \\
                            & = 2q\Omega_{0}^2y\boldsymbol{e_y}.
  \end{align}
We take the value $q = 1$ \citep[flat rotation curve][]{Ostriker2002}. Since $\Omega(R_0) \propto 1/R_0$, we also notice that $V_{shear}=-\Omega'(R_0)R_0 =\Omega(R_0)$.
From \citet{Binney1987}, $\Omega(R_0)=28 \ \mathrm{km.s^{-1}.kpc^{-1}}$ corresponds to the value in the solar neighbourhood.

To this effective potential, we must add the potential from self-gravity. 
We also add an analytical gravity profile accounting
for the distribution of stars and dark matter \citep{Kuijken89b}:
  \begin{equation}
\phi_{ext}(z) = K \left(\sqrt{z^2 + z_0^2} - z_0\right) + Fz^2,
\label{eq:phi_ana}
  \end{equation}
with $K = 1.42 \times 10^{-3}\ \mathrm{kpc}\ \mathrm{Myr}^{-2}$, $F = 2.75 \times 10^{-4}\ \mathrm{Myr}^{-2}$ , and $z_0 = 180\ \mathrm{pc}$ \citep{Joung06}.
Through a Poisson equation, the potential is associated with a density,
  \begin{equation}
\rho_{ext} = \frac{K}{4\pi G}\frac{D^2}{(z^2+D^2)^{3/2}} + \frac{F}{2\pi G}
\label{eq:rho_ana}
  .\end{equation}

Adding the Coriolis force, the MHD equations that are solved are
  \begin{align}
\partial_t \rho + \vec{\nabla} \cdot \left( \rho \vec{v} \right) &= 0, \\
\partial_t \left( \rho \vec{v} \right)
  + \vec{\nabla} \cdot \left( \rho \vec{v} \otimes \vec{v}
  + \left( P + \frac{B^2}{8\pi} \right) \tens{I}
  - \frac{\vec{B} \otimes \vec{B}}{4\pi} \right)
             &= -\rho\vec{\nabla} \Phi \nonumber \\
             & + 2q\rho\Omega_{0}^2y\,\vec{e_y} \nonumber \\
             & - 2\rho\vec{\Omega_0}\times\vec{v}, \label{eq:mhd-momentum}\\
\partial_t E
  + \vec{\nabla} \cdot \left(\left( E + P + \frac{B^2}{8\pi} \right) \vec{v}
  - \frac{1}{4\pi} \left( \vec{v} \cdot \vec{B} \right) \vec{B} \right)
             &= -\rho \vec{v} \cdot \vec{\nabla} \Phi - \rho\mathcal{L}, \\
\partial_t \vec{B} + \vec{\nabla} \cdot \left( \vec{v} \otimes \vec{B}
  - \vec{B} \otimes \vec{v} \right) &= 0, \\
\Delta \phi - 4\pi G \rho &= 0,\label{eq:mhd-poisson}
  \end{align}
with $\rho$, $\vec{v}$, $P$, $\vec{B}$, $\Phi$, and $E$ respectively being
the mass density, velocity, pressure, magnetic field, total gravitational
potential, and total (kinetic plus thermal plus magnetic) energy. \newline
The total gravitational potential is written 
$\Phi = \phi + \phi_{ext}$, with $\phi_{ext}$ being the analytical potential as explained above. For non-ionised gas, the loss function $\mathcal{L}$ is such that $n^2 \mathcal{L} = n^2 \Lambda(T) - n\Gamma$, where $n$ is the particle density, $\Gamma$ represents 
constant and uniform UV heating, and $\Lambda(T)$ is the cooling function, similar to the one used by \citet{Joung06}. For photoionised gas (HII regions), radiative cooling and heating is performed by the radiative transfer module RAMSES-RT (\citet{Rosdahl2013}).
We use the same model as in \citet{geen2016}, where they add radiative heating and cooling functions for metals.
\subsection{Boundary conditions and shearing box}
\label{sec:bcshbox}
\begin{figure}
    \begin{center}
      \includegraphics[width=0.45\textwidth]{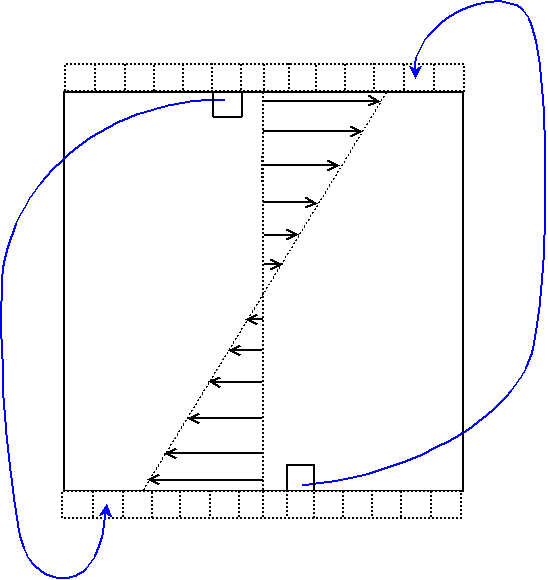}
    \end{center}
    \caption{Drawing of the shearing box used at the $y$ boundaries.}
    \label{fig:shearingbox}
\end{figure}
A schematic of the simulated volume in the context of the galactic disk, with axis directions, is shown in Fig. \ref{fig:shear}. 
We use periodic boundary conditions along the $x$-axis, and zero-gradient (except for $\Phi$) boundary conditions
along the $z$-axis (outside the galactic plane).
The most difficult part concerns the boundary conditions along the $y$-axis. Simple periodic
conditions cannot be used, as the two boundaries have opposite and non-zero velocities in our rotating frame (see Sect. \ref{sec:eq}). 
A shearing box is therefore implemented \citet{Hawley95} (see Fig. \ref{fig:shearingbox}).\\
In each ghost cell in the upper $y$ boundary, the values of $\rho$, $v_y$, $v_z$ , and $T$ are equal to those of
the corresponding cell in the lower domain, shifted of $-V_{shear} \cdot t$ in the $x$-direction.
For the $x$-velocity $v_x$, it is not simply the value from the lower cell, but
the necessary differential of velocity that is added: $v_x~+~V_{shear}$ is applied in the ghost cell.

The treatment of the magnetic field is a bit different because
zero-divergence must be assured in the cells, as well as the continuity of the normal field at the boundary
(in RAMSES, each cell has six values for the magnetic field, one on each face).
We do the following: first the values of the tangential fields
($B_x$ and $B_z$) are applied in the ghost cell. They are read as usual in the corresponding cell of the domain. The continuity
of $\vec{B}$ is then assured by taking the value of the normal $B_y$ of the cell just below.
Finally, the remaining unknown $B_y$ in the ghost cell is specifically chosen in order to ensure 
the zero-divergence of the magnetic field. The same treatment is applied in the $z$-boundaries, adapted to zero-gradient conditions.
Then, the energy per unit of mass must be defined in the ghost cell:  the internal energy from the cell
of the domain is taken, but the kinetic and magnetic energies have to be re-calculated.

The last physicial parameter that remains unknown in the ghost cell 
is the gravitational potential $\Phi$. The same treatment cannot be applied,
since the Poisson solver is an iterative solver (for a single timestep), but RAMSES 
does not update the boundary conditions at each iteration of the Poisson solver. 
Therefore, an analytical boundary condition for $\Phi$ has to be taken (at the $y$-boundaries and $z$-boundaries 
only, the $x$-boundaries being periodic and correctly solved), 
which does not need to be updated at each of these iterations (but that is updated at each timestep). \\
We proceed as follows: we notice that the external potential $\phi_{ext}$ (Eq. \ref{eq:phi_ana}) 
is far stronger than the self-gravitating potential $\phi$.
Hence, in order to find the analytical boundary condition, 
an approximation is made that the self-gravitating potential in the box is of the form 
\begin{equation}
        \phi_{app} (z)=\frac{4\pi G\alpha}{z_0^2}\left(\sqrt{z^2+z_0^2}-z_0\right)
,\end{equation}
with $\alpha$ and $z_0$ to be determined at each timestep.
This approximate $\phi_{app}$ will be used in the boundary condition 
\begin{equation}
\Phi_{bound}~=~\phi_{app}~+~\phi_{ext}
.\end{equation}
Via the Poisson equation, this potential is associated with an approximate density
\begin{equation}
        \rho_{app} (z)=\frac{\alpha}{\left(z^2+z_0^2\right)^{3/2}}
.\end{equation}
By integrating this density, $\alpha$ is deduced from the total mass of gas $M$ in our box,
\begin{equation}
  M=\alpha \frac{L^3}{z_0^2\sqrt{\frac{L^2}{4}+z_0^2}}
.\end{equation}
To estimate the scale height, $z_0$, we simply find the altitude at which the density is given by
\begin{equation}
  \frac{\rho_{app}(z_0)}{\rho_{app}(0)} = \frac{1}{2^{3/2}} \simeq \frac{M_{gas}(z_0)}{M_{gas, max}}
.\end{equation}
Finally, the boundary condition used in the $y$ and $z$ directions is
\begin{equation}
        \Phi_{bound}=\phi_{ext}+\frac{4\pi GM}{L^3}\sqrt{\frac{L^2}{4}+z_0^2} \, \left(\sqrt{z^2+z_0^2}-z_0\right)
  \label{eq:phi_bound}
.\end{equation}
With this boundary condition, the gas is stable on large scales while otherwise a global collapse occurs if the usual shearing box treatment is applied.
We have performed a test for the accuracy of these gravitational boundaries (see Appendix \ref{ap:bc})
and the error is around $30-40 \%$, which is lower than the incertitude coming from our numerical setup (see Sect. \ref{sec:num}).

The third section of \citet{Shen2006} describes hydrodynamic tests for shearing boxes. They have been performed in our case,
and are in good agreement with analytical results.
\subsection{Initial conditions}

The size of our cubic box is $1 \ \mathrm{kpc}$.
We impose initially a Gaussian density profile (for hydrogen atoms) along the $z-axis$,
  \begin{equation}
n(z) = n_0 \exp \left[ - \frac{1}{2} \left( \frac{z}{z_0}  \right)^2 \right],
  \end{equation}
with $n_0 = 1.5\ \mathrm{cm^{-3}}$ and $z_0 = 150\ \mathrm{pc}$.
The initial column density is then $\Sigma_{gas} = \sqrt{2\pi}m_pn_0z_0 = 4 \cdot 10^{-3} \ \mathrm{g.cm^{-2}} = 19.1 \ \mathrm{M_{\odot} \, pc^{-2}}$
, where $m_p=1.4 \times 1.66~\cdot~10^{-24} \ \mathrm{g}$ is the mean mass per hydrogen atom.
The initial temperature is chosen to be $8000 \ \mathrm{K}$. This is a typical temperature of the warm neutral medium (WNM) phase of the ISM. We also add an initial turbulent velocity field with a root mean square (RMS) dispersion of $5 \ \mathrm{km / s}$ 
and a Kolmogorov power spectrum with random phase \citep{Kolmogorov41}.

Finally, we add a Gaussian magnetic field, oriented along the $x-axis$,
  \begin{equation}
B_x(z) = B_0 \exp \left[ - \frac{1}{2} \left( \frac{z}{z_0} \right)^2 \right],
  \end{equation}
with $B_0 \simeq 0 \text{ or } 4\ \mathrm{\mu G}$ depending on the simulation (magnetohydrodynamic or just hydrodynamic). \\
Our different runs are detailed in Table \ref{tab:runs}.
  \begin{table*}
    \begin{center}
      \begin{tabular}{lccc}
        \hline\hline
        Name & $B_0 \ (\mathrm{\mu G}) $ & $V_{shear} \ (\mathrm{km.s^{-1}.kpc^{-1}})$ & Particularities \\
        \hline
        noSH\_hydro & 0 & 0  & -- \\
        SH14\_hydro & 0 & $14$ & -- \\
        SH28\_hydro & 0 & $28$ & -- \\
        SH28\_hydro\_VSAT\_TSAT & 0 & $28$ & Increased $T_{sat}$ and $V_{sat}$ \\
        SH56\_hydro & 0 & $56$ & -- \\
        \hline
        noSH\_mhd & $4$ & 0  & -- \\
        SH14\_mhd & $4$ & $14$ & -- \\
        SH28\_mhd & $4$ & $28$ & -- \\
        SH28\_mhd\_VSAT\_TSAT & $4$ & $28$ & Increased $T_{sat}$ and $V_{sat}$ \\
        SH28\_mhd\_vdisp & $4$ & $28$ & The massive stars are motionless around the sinks \\
        SH28\_mhd\_lowsinkthres & $4$ & $28$ & The creation and accretion threshold of the sinks is 4x lower \\
        SH28\_mhd\_highsinkthres & $4$ & $28$ & The creation and accretion threshold of the sinks is 4x higher \\
        SH28\_mhd\_nosn & $4$ & $28$ & No supernovae feedback \\
        SH28\_mhd\_nort & $4$ & $28$ & No HII regions feedback \\
        SH28\_mhd\_nort\_nosn & $4$ & $28$ & No HII and supernovae feedback \\
        SH56\_mhd & $4$ & $56$ & -- \\
        \hline
        SH28\_mhd\_512 & $4$ & $28$ & Resolution $512^3$ \\
        SH56\_mhd\_512 & $4$ & $56$ & Resolution $512^3$ \\
        \hline
        noSH\_mhd\_novir & $4$ & $0$ & No virial tests for the sink particles (see Sect. \ref{sec:sink}) \\
        noSH\_mhd\_novir\_trueperiodic & $4$ & $0$ & No virial tests, true periodic boundary conditions for gravity\\
        SH28\_mhd\_novir & $4$ & $28$ & No virial tests \\
        SH28\_mhd\_novir\_lowsinkthres & $4$ & $28$ & No virial tests and the threshold of the sinks is 4x lower \\
        SH28\_mhd\_novir\_highsinkthres & $4$ & $28$ & No virial tests and the threshold of the sinks is 4x higher \\
        SH56\_mhd\_novir & $4$ & $56$ & No virial tests \\
        \hline
      \end{tabular}
    \end{center}
    \captionsetup{justification=centering}
    \caption{Table of the different runs. Section \ref{sec:sat} gives the meaning of $T_{sat}$ and $V_{sat}$. 
             The case $V_{shear} = 28 \ \mathrm{km.s^{-1}.kpc^{-1}}$ corresponds to the value near the Sun (see Sect. \ref{sec:sfr}).}
    \label{tab:runs}
  \end{table*}  

The Toomre criterion for our shearing disk of gas is \citep{Binney1987}
\begin{equation}
Q = \frac{\sqrt{2} c_s \Omega}{\pi G \Sigma_{gas}}
\label{eq:Toomre}
,\end{equation}
where $c_s$ is the sound speed of the gas.
This disk is stable to axisymmetric perturbations if
\begin{equation}
Q > 1
\label{eq:stab}
.\end{equation}
With a speed of sound of $8 \ \mathrm{km.s^{-1}}$, the values for our simulations are listed in Table \ref{tab:Toomre}.
These values are very close to the stability criterion, hence the consistency of our parameters.


\subsection{Sink particles}
\label{sec:sink}

The star-forming regions are represented by Lagrangian sink particles \citep{Krumholz04, Bleuler2014}:
the resolution being too low to describe each star individually, they are described as a group
by subgrid models. The sink particles are created when a density threshold is reached (usually $10^3 \ \mathrm{cm^{-3}}$, see Sect. \ref{sec:num}), and 
when virial criteria are satisfied (see \citet{Bleuler2014}). As the particles represent a group of stars and because the resolution is coarse,
it is unclear that virial criteria should be used. Therefore we have performed simulations with and without the criteria.
In the following, unless explicitly mentioned, the virial criteria are being tested.
The sink particles gain mass by accreting gas from their surroundings: when the density at the position of the sink is higher than this same threshold, 
the mass excess is accreted.\\

\subsection{Supernova and HII feedback}

\label{sec:sat}

\begin{figure}
  \begin{center}
    \includegraphics[width=0.5\textwidth]{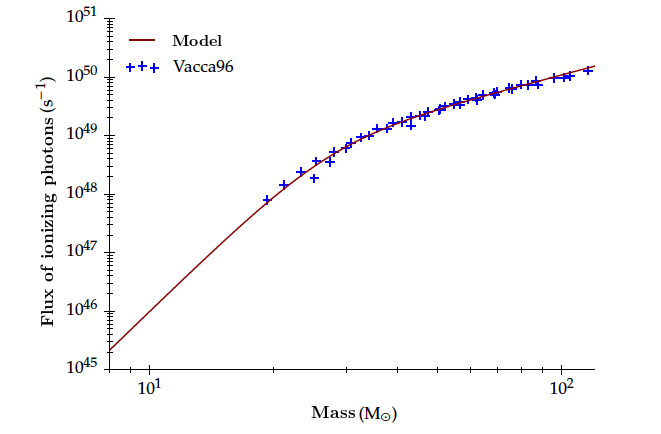}
  \end{center}
  \caption{Fitting of Eq. \ref{eq:flux} with the data of \citet{Vacca1996}.}
  \label{fig:fit}
\end{figure}
The supernova feedback implemented in RAMSES is based on scheme C of \citet{Hennebelle14}:
a massive star is formed each time a sink particle accretes more than $120\ \mathrm{M_\odot}$.
Its mass is determined randomly assuming a Salpeter initial mass function \citep{Salpeter55, Chabrier2003}.
This massive star is placed randomly within a $10 \ \mathrm{pc}$ radius around the sink.
Its lifetime is calculated from its mass, and  the supernova is triggered 
at the end of its life at its current position: we assume the star has moved
with a velocity of $v_{disp} = 1 \ \mathrm{km/s}$ in the sink frame throughout its life.
Simulations with a lower or higher density threshold, or with $v_{disp} = 0$ have also been performed,
and are discussed in Sect. \ref{sec:num}. The supernova then consists of an injection of $4 \cdot 10^{43}\ \mathrm{g\ cm / s}$ 
\citep{Iffrig15a} of radial momentum around the place of the star.

We manually limit the temperature to $T_{sat}~=~10^6\ \mathrm{K}$ 
and the velocity provided by the supernova to  $V_{sat}~=~500\ \mathrm{km/s}$ 
in order to prevent too high sound speeds and too small timesteps.
We also performed some simulations with greater values (see Table \ref{tab:runs}): 
$T_{sat}~=~10^7\ \mathrm{K}$ and $V_{sat}~=~1000\ \mathrm{km/s}$, to analyse their influence. \\
The maximum of velocity $V_{sat}$ is set higher than in the previous simulations because now,
as previously explained, the supernova is not set immediately 
but rather at the death of the star: therefore, the explosion may happen in diffuse 
gas and the maximum velocity would be too easily reached, thus causing a loss of precision.

Feedback by HII regions is also implemented in the code. The energy and momentum injected by the HII region (via the pressure gradient) in the interstellar medium is highly dependent on
the flux of ionising photons $S_*$ emitted by the star. Using the data of \citet{Vacca1996},
we predicate the following relation between $S_*$ and the mass of the star:
\begin{equation}
        {S\!}_* = 2^{\beta}{S\!}_0\frac{\left(\frac{M}{M_0}\right)^{\alpha}}{\left(1+\left(\frac{M}{M_0}\right)^{\frac{\alpha-\gamma}{\beta}}\right)^{\beta}}
\label{eq:flux}
.\end{equation}
We then have the asymptotic relations
\begin{equation}
        {S\!}_* \sim
    \begin{cases}
            2^{\beta}{S\!}_0\left(\frac{M}{M_0}\right)^{\alpha} & \mathrm{if} \ M \ll M_0, \\
            2^{\beta}{S\!}_0\left(\frac{M}{M_0}\right)^{\gamma} & \mathrm{if} \ M \gg M_0.
    \end{cases}
.\end{equation}
The results of this empirical model are given in Table \ref{tab:fit} and Fig. \ref{fig:fit}.
The evolution of HII regions is then computed with the radiative transfer module RAMSES-RT \citep{Rosdahl2013}.
Comparisons between analytical models and simulations results for the expansion of HII regions with
RAMSES-RT have been done by \citet{Geen2015}.
RAMSES-RT has been validated by the STARBENCH comparison project (\citet{Bisbas15}), 
which tested the D-type expansion of an HII region with different codes.
Other works have recently studied the role of HII regions in star formation \citep{Butler2017, Peters2017}.

\begin{table}
  \begin{center}
    \begin{tabular}{cc}
      \hline\hline
      $V_{shear} \ (\mathrm{km.s^{-1}.kpc^{-1}})$ & Q \\
      \hline
      14 & 0.61 \\
      28 & 1.2 \\
      56 & 2.4 \\
      \hline
    \end{tabular}
  \end{center}
  \caption{Values of the Toomre criteria (Eq. \ref{eq:Toomre}) for different values of $V_{shear}$.}
  \label{tab:Toomre}
\end{table}
\begin{table}
  \begin{center}
    \begin{tabular}{ccc}
    \hline \hline
    Parameter & Value & Units \\
    \hline
    $S_0$ & $4.36525 \times 10^{48}$ & $\mathrm{s}^{-1}$ \\
    $M_0$ & $27.2810$ & $\mathrm{M}_{\odot}$ \\
    $\alpha$ & $6.84002$ \\
    $\beta$ & $1.14217$ \\
    $\gamma$ & $1.86746$ \\
    \hline
    \end{tabular}
  \end{center}
  \caption{Parameters of  Eq. (\ref{eq:flux}).}
  \label{tab:fit}
\end{table}

\section{Results}
\subsection{Global considerations}
\label{sec:shbox}

\begin{figure*}
  \begin{center}
    \includegraphics[width=0.83\textwidth]{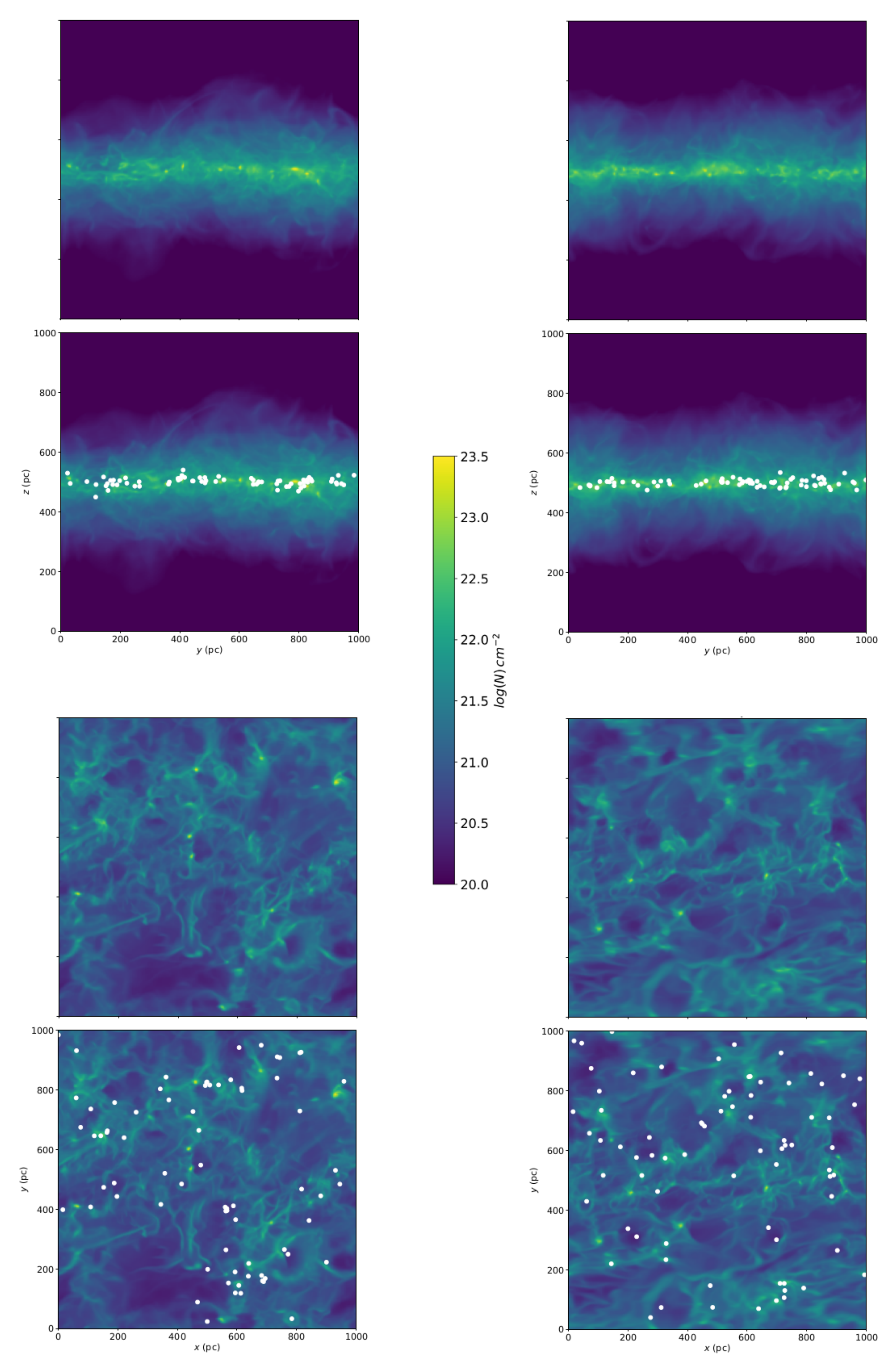}
  \end{center}
  \caption{Column density for our MHD simulations ($256^3$), with feedback by supernovae and HII regions, at $80 \ \mathrm{Myrs}$. 
          For each plane, there are two rows: the first one does not display sink particles,
          but the second one does, each one representing a star cluster (the colours are random).
          \textbf{Left:} $V_{shear}=28 \ \mathrm{km.s^{-1}.kpc^{-1}}$. \textbf{Right:} $V_{shear}=56 \ \mathrm{km.s^{-1}.kpc^{-1}}$.}
  \label{fig:mhd}
\end{figure*}

\begin{figure*}
  \begin{center}
    \includegraphics[width=\textwidth]{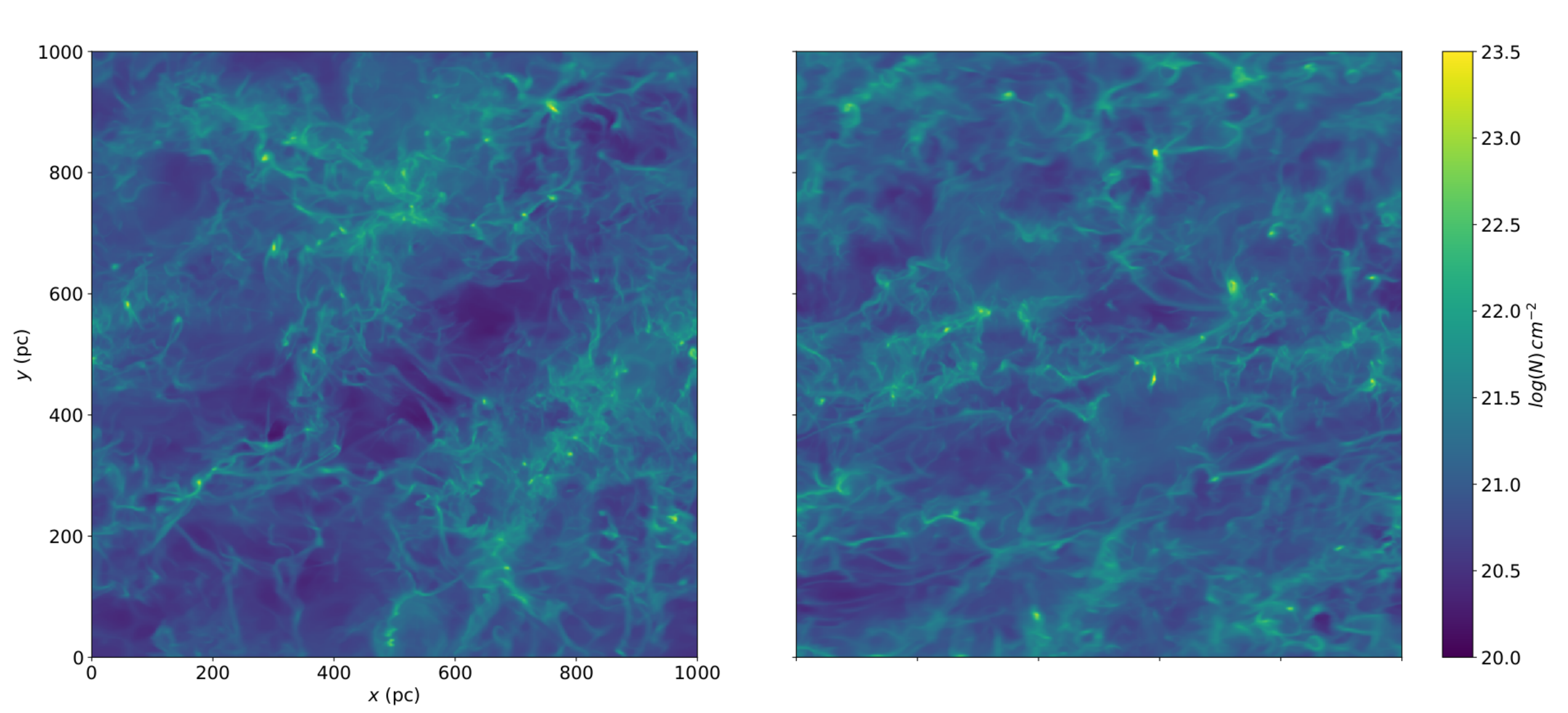}
  \end{center}
  \caption{Column density for our MHD simulations ($512^3$), with feedback by supernovae and HII regions, at $80 \ \mathrm{Myrs}$. \\
            \textbf{Left:} $V_{shear}=28 \ \mathrm{km.s^{-1}.kpc^{-1}}$. \textbf{Right:} $V_{shear}=56 \ \mathrm{km.s^{-1}.kpc^{-1}}$. }
  \label{fig:mhd_512}
\end{figure*}

Figure \ref{fig:mhd} shows the column density in two of the MHD simulations (at a resolution of $256^3$), with
feedback by supernovae and HII regions, at $80 \ \mathrm{Myrs}$. 
The coloured points are sink particles, as explained in Sect. \ref{sec:sink}. 
We can see the effects of the shearing box in these images: the $y-boundaries$ are now well defined, with gas and
sink particles getting through (with a corresponding new position and velocity).
Figure \ref{fig:mhd_512} shows the same physical simulations at the same time, at a higher resolution. The global patterns are thus broadly the same,
but the small scales and the small high density regions become more visible.
The impact of the shear is that the gas is more scattered at higher shear, and more condensed at lower shear.
We notice that the galactic shear tends to stretch the large structures of gas: it competes with gravity and opposes
the collapse of the molecular clouds. Hence, the galactic shear is expected to reduce the star formation rate.
We will quantify this effect in Sects. \ref{sec:sfr} and \ref{sec:structures}.

\begin{figure*}
  \begin{center}
    \includegraphics[width=0.95\textwidth]{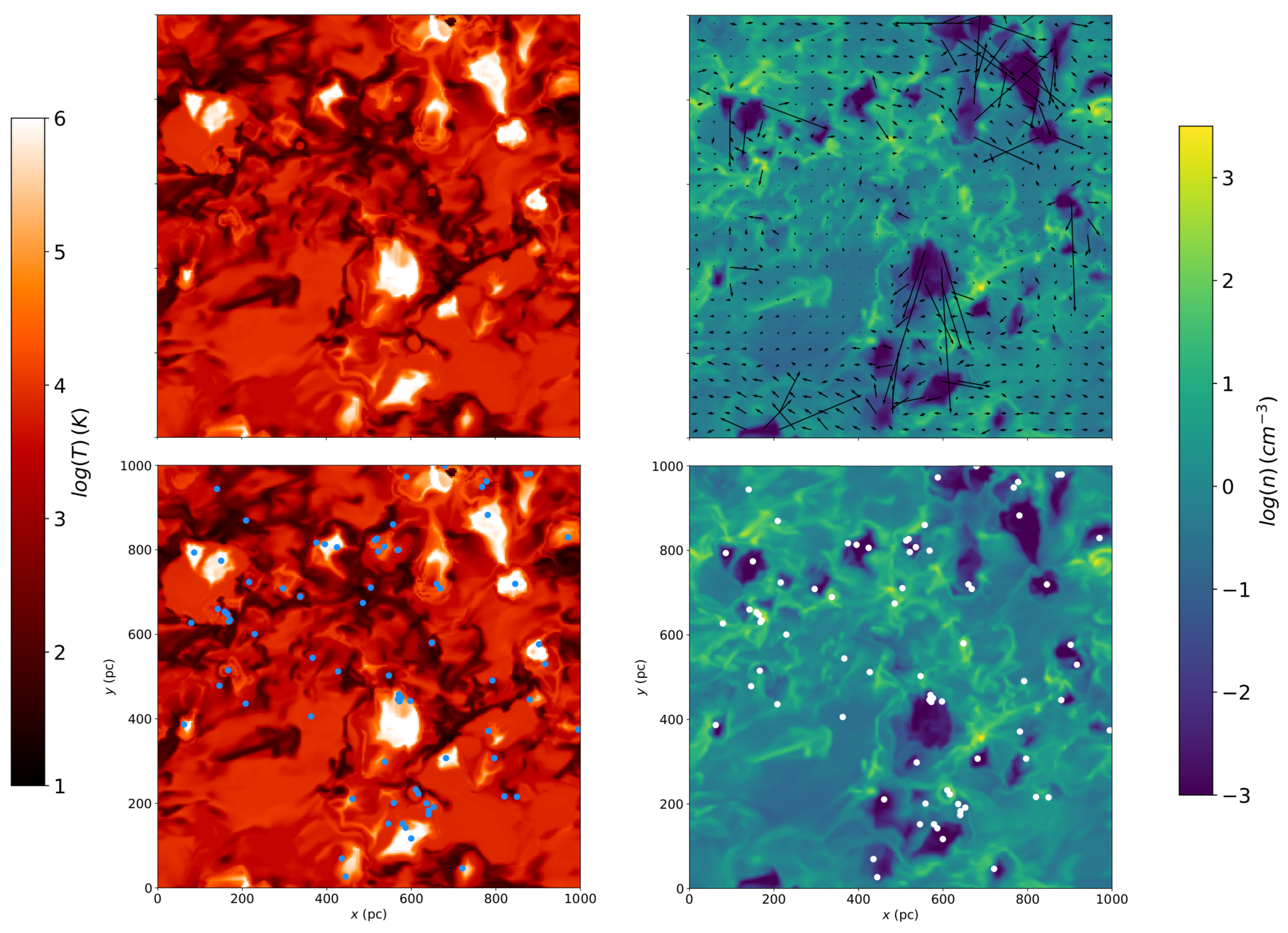}
  \end{center}
  \caption{Density and temperature maps of our SH28\_mhd simulation ($256^3$), at $85 \ \mathrm{Myrs}$. The arrows on
          the density map represent the velocity of the gas. The regions above $10^5 \ K$ are due to the supernovae.}
  \label{fig:sn}
\end{figure*}

Supernovae and HII regions are also implemented, as explained in Sect. \ref{sec:sat}. The temperature and density maps of Fig.~\ref{fig:sn}
show the effects of supernovae on the interstellar gas: there are peaks of temperature and minima of density at the supernovae locations.
The surrounding gas is heated and dispersed (see Sect. \ref{sec:sfr} for the effect on star formation).
Some of the explosions are at the sink particle location, but some others are lightly shifted from the star cluster.
This is the consequence of the scheme described in Sect. \ref{sec:sat}: the star is moving, and the supernovae is triggered
at the end of its life.


\subsection{Influence of the physical parameters on the star formation rate (SFR)}
\label{sec:sfr}

\begin{figure*}
  \begin{subfigure}[]{\columnwidth}
    \centering
    \includegraphics[width=7.1cm]{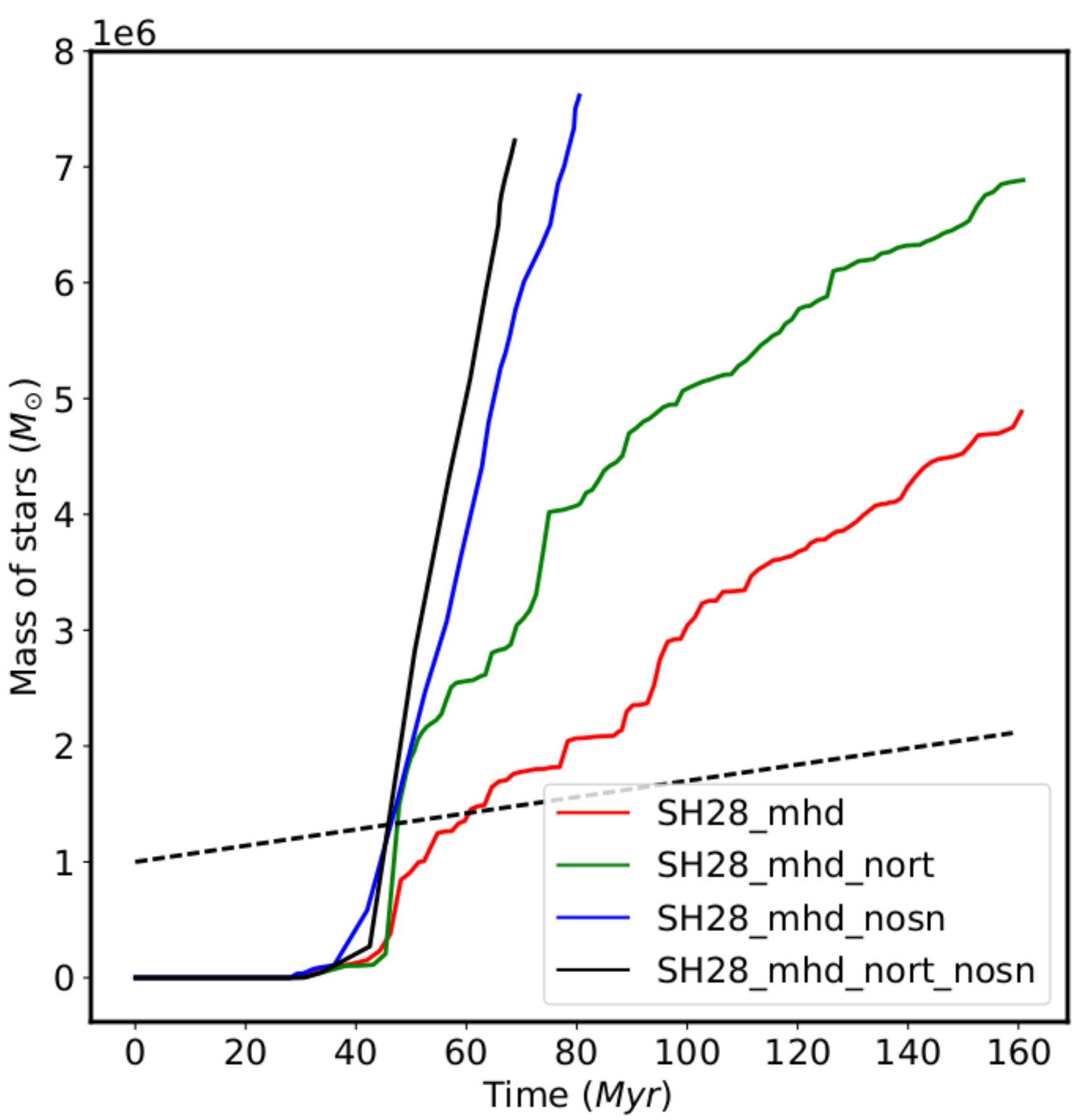}
    \caption{}
  \end{subfigure}
  \begin{subfigure}[]{\columnwidth}
    \centering
    \includegraphics[width=7.1cm]{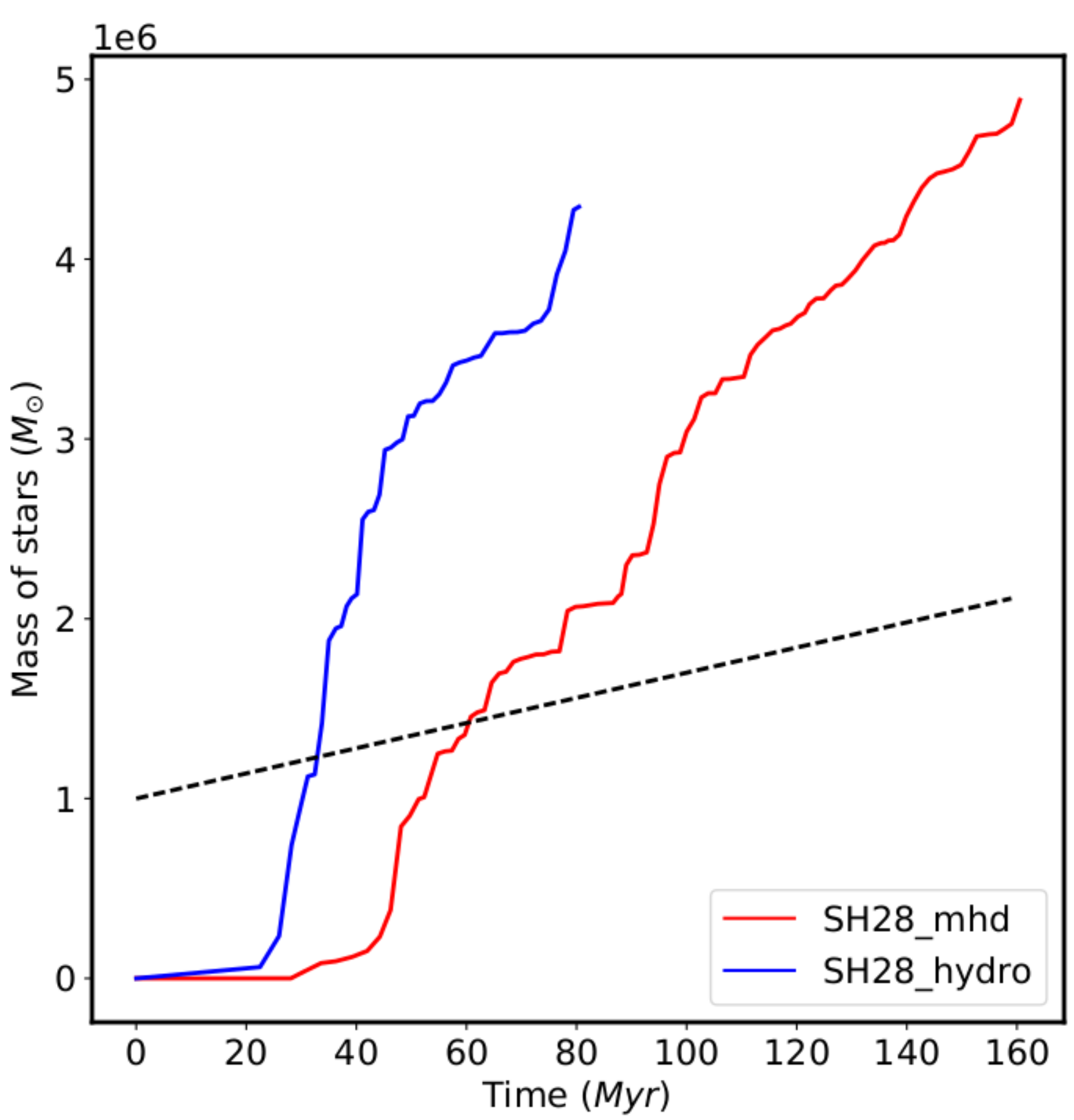}
    \caption{}
  \end{subfigure}
  \begin{subfigure}[]{\columnwidth}
    \centering
    \includegraphics[width=7.1cm]{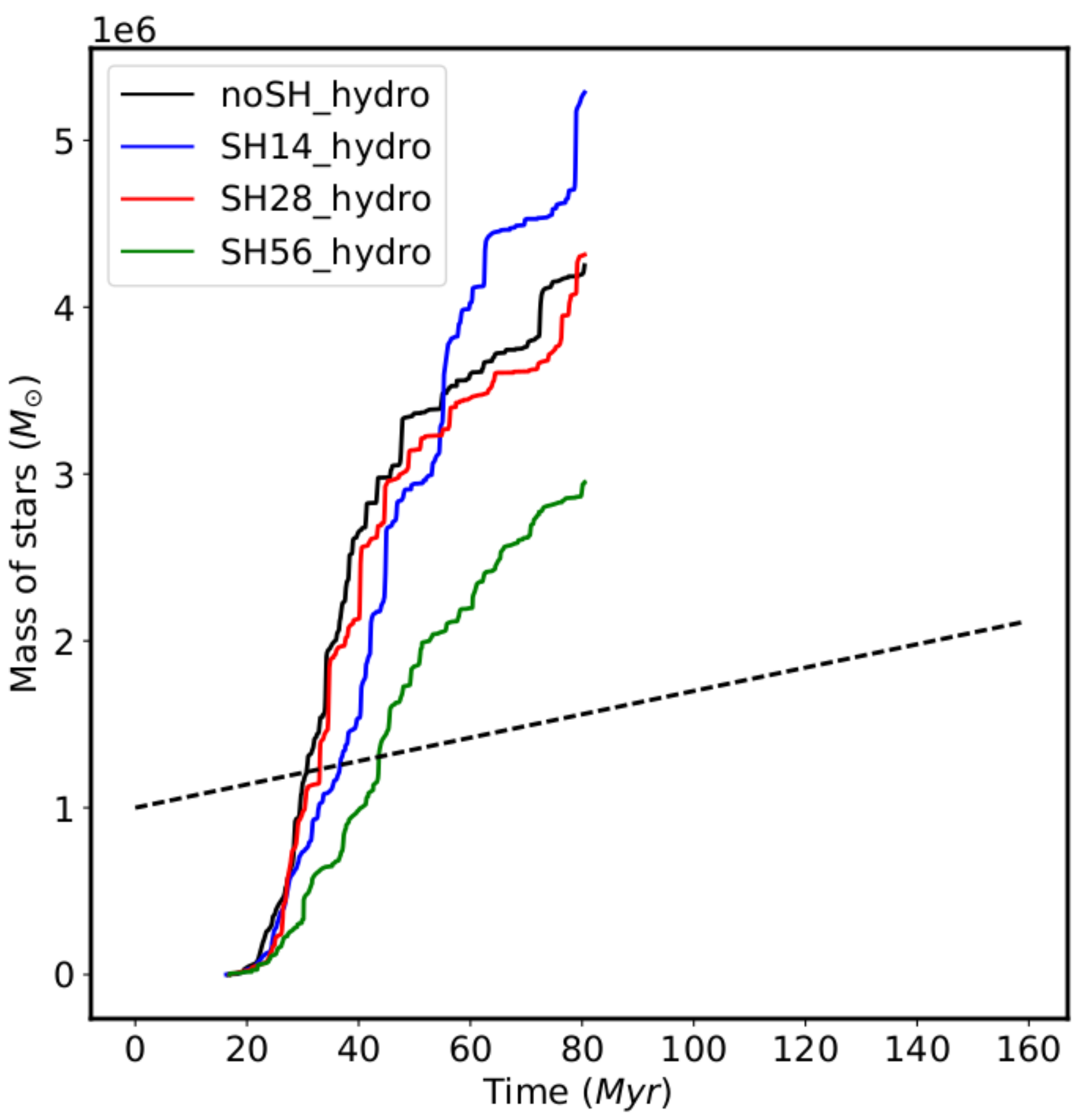}
    \caption{}
  \end{subfigure}
  \centering
  \begin{subfigure}[]{\columnwidth}
    \centering
    \includegraphics[width=7.1cm]{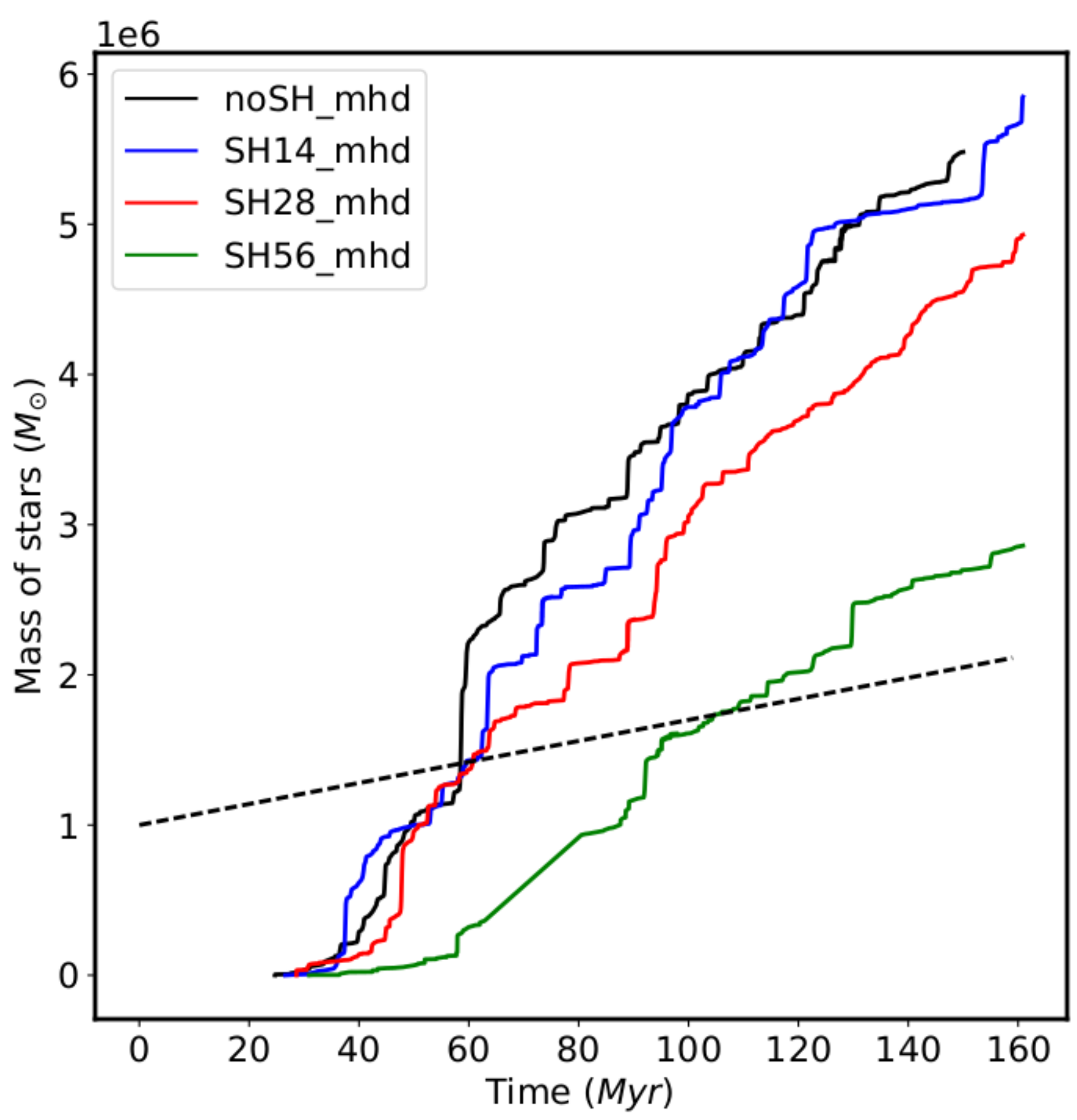}
    \caption{}
  \end{subfigure}
  \caption{Total mass of stars formed over time for different simulations with different \textbf{physical} parameters.
             The slope of the dashed line is the observed SFR for our $\Sigma_{gas}$ \citep{Kennicutt12}. 
Top-left panel shows the importance of the two types of feedback. Top-right panel shows the importance of 
magnetic field. Bottom panels illustrate the importance of the shear. Altogether we see that there is no dominant effect
but instead all physical processes, namely feedback by supernovae, ionising radiation, magnetic field, and shear, all contribute
appreciably to reduce the SFR.}
  \label{fig:plot_sfr1}
\end{figure*}

Figure \ref{fig:plot_sfr1} displays the total mass of stars (in units of $M_{\odot}$) 
formed over time in our $1 \ \mathrm{kpc}^3$ box, for different physical simulations. 
The \textbf{slope} of the dashed line is the observed SFR given our column density, which we want to reproduce. 
This value $\Sigma_{SFR} \simeq 7.10^{-3} \ M_{\odot}.\mathrm{yr^{-1}.kpc^{-2}}$ comes from Fig. 11 of \citet{Kennicutt12}.

$\bullet$ \textbf{Panel (a)} shows the influence of the feedback (by supernovae and HII regions) on the total mass of stars.
Without any kind of feedback (black curve), the mass of stars (hence the SFR) is very high.
When feedback by HII regions is added (blue curve), it decreases by $20\%$. 
The feedback by supernovae is, however, dominant: with supernovae and no HII regions (green curve), the asymptotic SFR decreases by a factor approximately three.
The red curve is obtained with both kinds of feedback: it is now far closer to the observed SFR, and it shows the huge
impact of the feedback on the ISM. However, in these simulations, the impact of the supernovae is lesser and the SFR is greater 
than in \citet{Iffrig2017}, because the massive stars now explode with some delay (see Sect. \ref{sec:sat}). 
The explosions can now happen outside the clouds, in more diffuse gas,
where they are less effective at suppressing star formation.

\begin{figure}
  \begin{center}
    \includegraphics[width=0.47\textwidth]{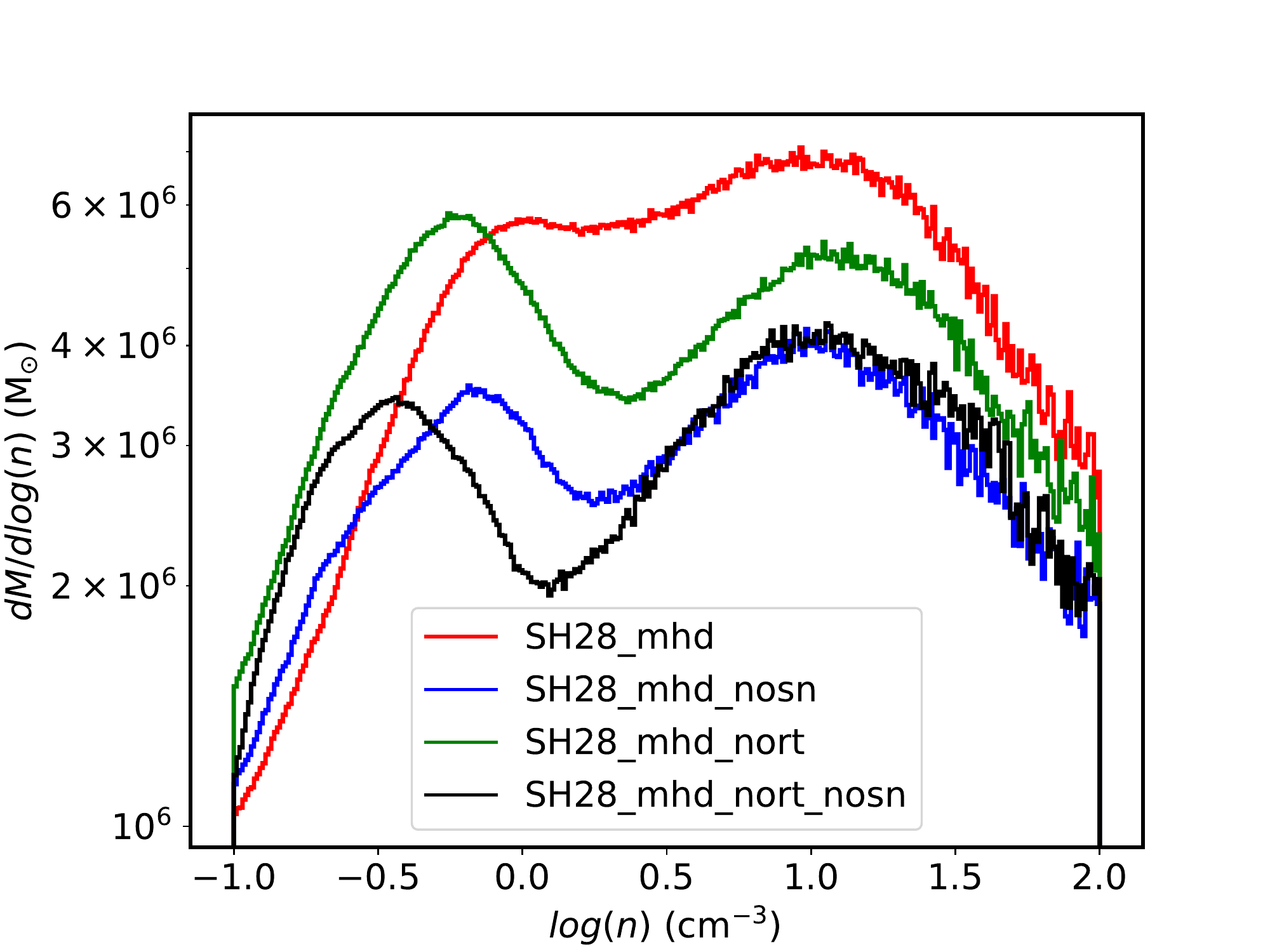}
  \end{center}
  \caption{Distribution functions of the density of the gas for SH28\_mhd simulations, with or without feedback, at $80$ Myrs.}
  \label{fig:pdf_feedback}
\end{figure}

There is a non-linear coupling between the supernovae and the HII regions 
that can be seen in the gas distributions of these four simulations, shown in Fig. \ref{fig:pdf_feedback}. 
When one kind of feedback is absent, there
is a drop in the distribution function, between $1$ and $2 \ \mathrm{cm^{-3}}$. Both kinds of feedback must be present for the hollow to disappear (red).
The effect of both supernovae and HII regions (on the SFR or on the PDF) is not the sum of the effects of the individual processes.
Such non-linear coupling was also noticed in the simulations of \citet{Bournaud16} where the outflow rate when both processes are present
was above the sum of the outflow rates in the individual cases.

$\bullet$ \textbf{Panel (b)} shows the influence of the magnetic field: without magnetic field, the total stellar mass increases by a factor of two.
The magnetic field still strongly suppresses the SFR, and has a strong impact on the shape of the ISM:
in the presence of magnetic field, the clumps are more filamentary than they are in pure hydrodynamic cases \citep{hennebelle2013}.
In MHD simulations, the matter is gathered in more stable filaments and that contributes to reduce the SFR.

$\bullet$ \textbf{Panels (c) and (d)}: These plots show how the total mass of stars depends on the shear.
The velocity gradient $V_{shear}$ takes the values: $0$ (black), $14$ (blue), $28$ (red), and $56$ (green) $\mathrm{km.s^{-1}.kpc^{-1}}$.
For the MHD cases (Panel (d)), the higher the gradient, the lower are the mass and the SFR. 
This is exactly the behaviour that was expected, as stated in Sect. \ref{sec:shbox}.\\
This is not so true for hydrodynamic simulations (Fig. (c)).
This comes from the fact that in the hydrodynamic case, the filaments are very unstable, and thus the shear is less effective.\\
In MHD, the case with $V_{shear}=14 \ \mathrm{km.s^{-1}.kpc^{-1}}$ (blue) is very similar to the one without shear (black).
This is probably because the Toomre stability criterion (\ref{eq:stab}) is not reached at this $V_{shear}$.\\
For the MHD simulations, the asymptotic behaviour of the green curve is close the observed SFR (too high by a factor of $\sim 1.5$).
Nevertheless, this is for a gradient $V_{shear} = 56 \ \mathrm{km.s^{-1}.kpc^{-1}}$, while the value in the Sun's neighbourhood
is $28 \ \mathrm{km.s^{-1}.kpc^{-1}}$ (see Sect. \ref{sec:eq}). 
As for the SFR for a velocity gradient $V_{shear} = 28 \ \mathrm{km.s^{-1}.kpc^{-1}}$ (red curve), it is still too high by a factor approximately four.
in the simulations compared to observations.
However, as described in Sect. \ref{sec:sh_vs_density}, $56 \ \mathrm{km.s^{-1}.kpc^{-1}}$ may be the right value to use. 


\subsection{Numerical parameters and numerical convergence}
\label{sec:num}

\begin{figure*}
  \begin{subfigure}[]{\columnwidth}
    \centering
    \includegraphics[width=6.9cm]{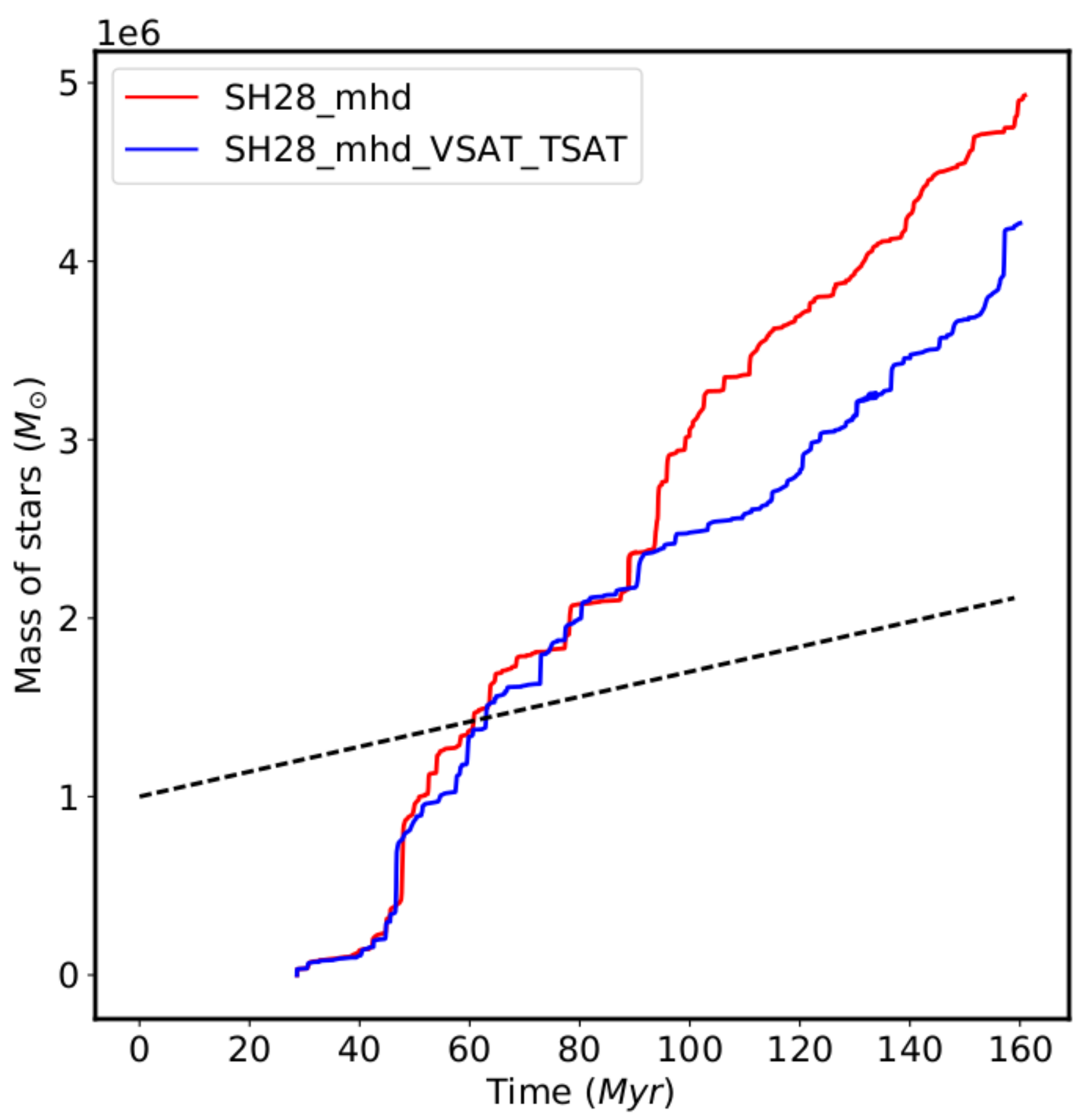}
    \caption{}
  \end{subfigure}
  \begin{subfigure}[]{\columnwidth}
    \centering
    \includegraphics[width=6.9cm]{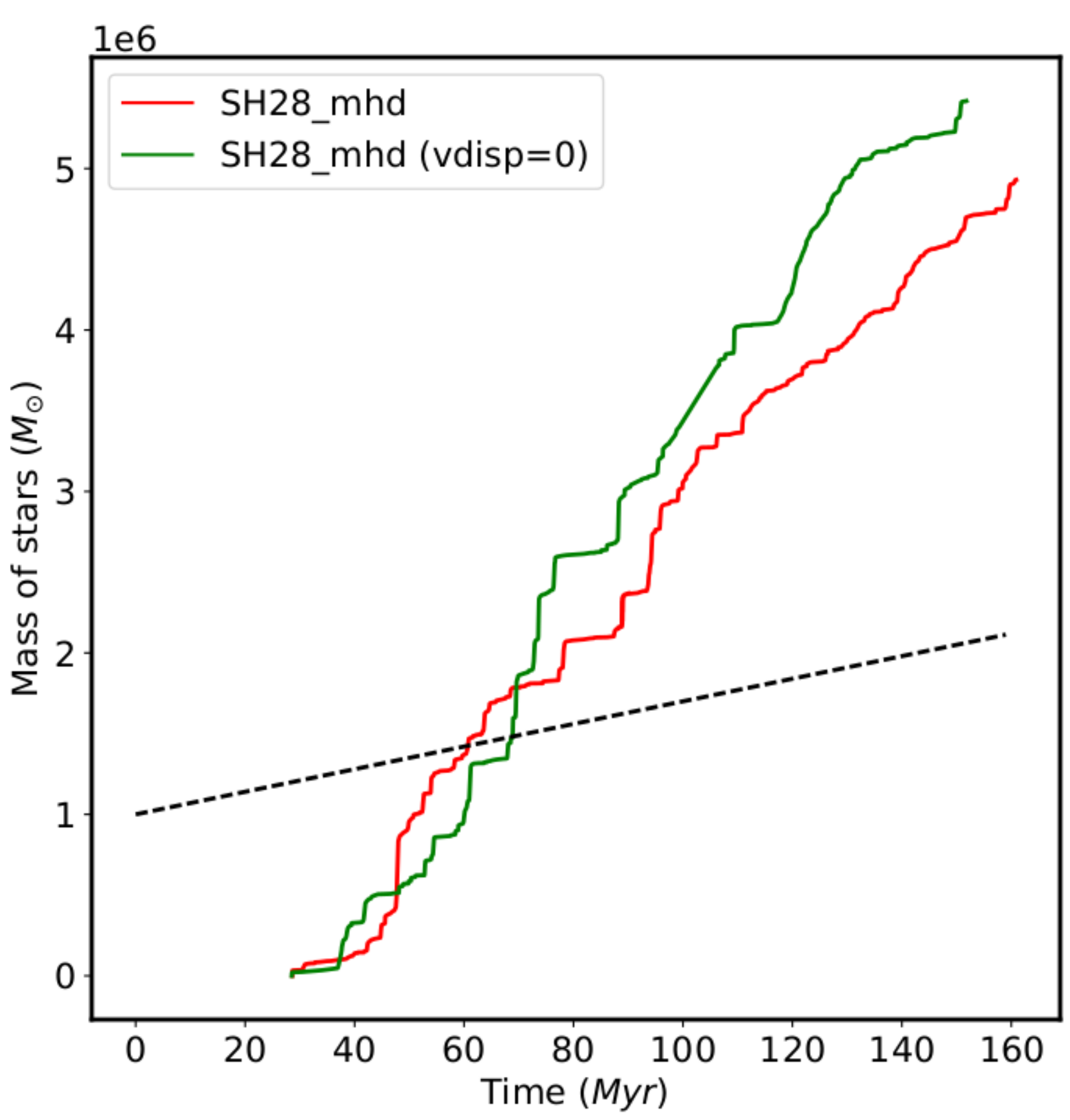}
    \caption{}
  \end{subfigure}
  \begin{subfigure}[]{\columnwidth}
    \centering
    \includegraphics[width=6.9cm]{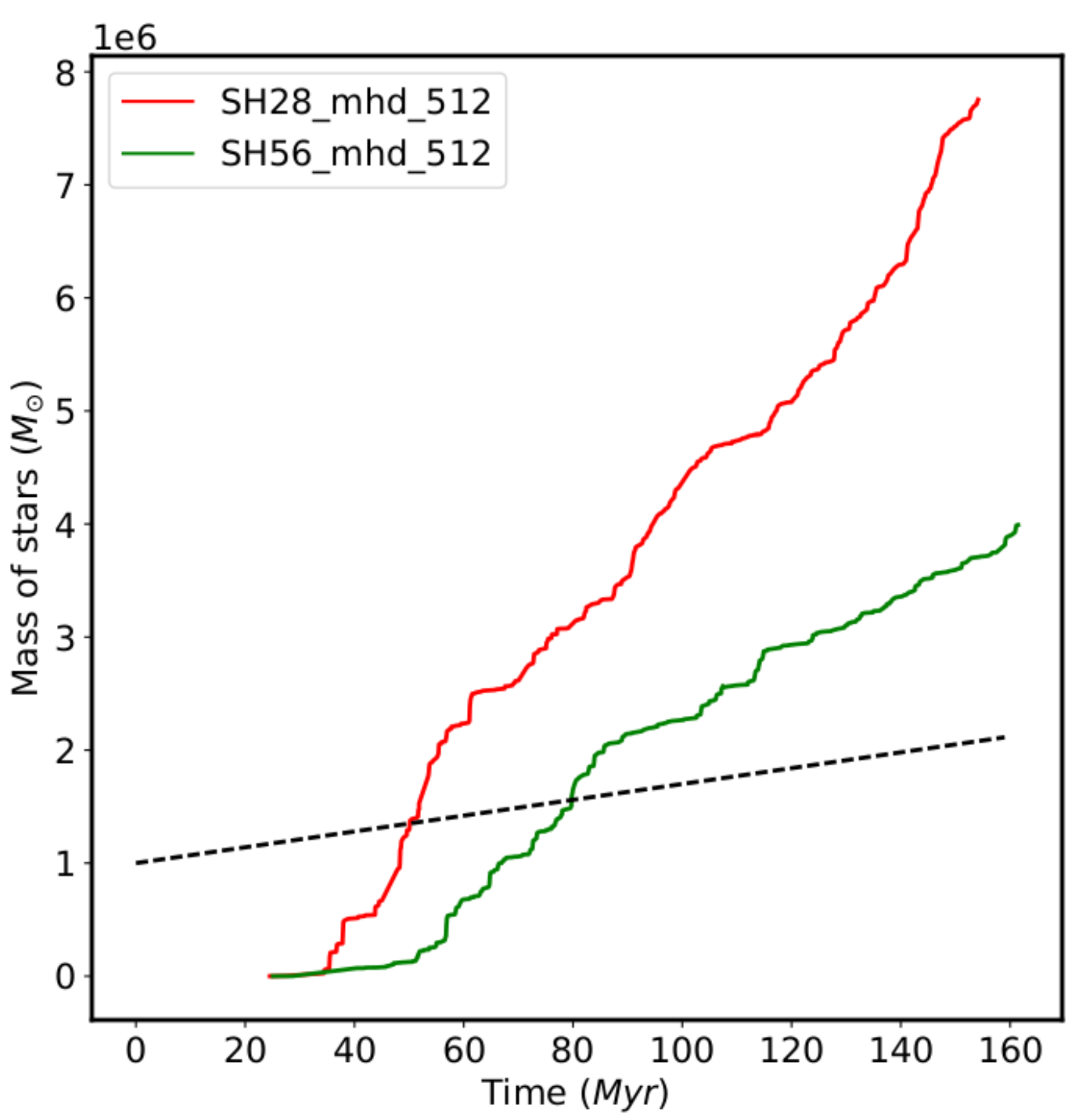}
    \caption{}
  \end{subfigure}
  \begin{subfigure}[]{\columnwidth}
    \centering
    \includegraphics[width=6.9cm]{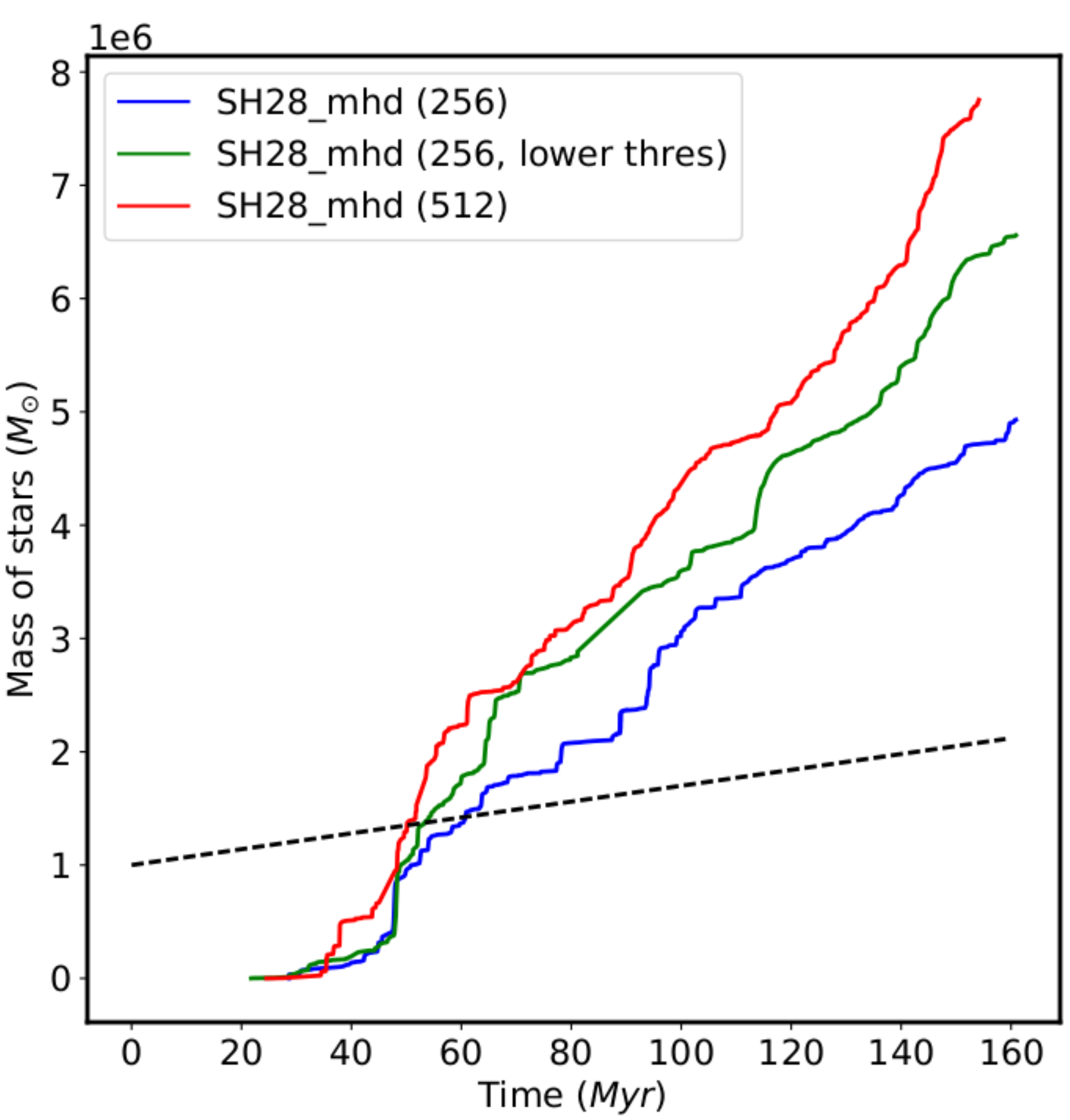}
    \caption{}
  \end{subfigure}
  \centering
  \begin{subfigure}[]{\columnwidth}
    \centering
\includegraphics[width=6.9cm]{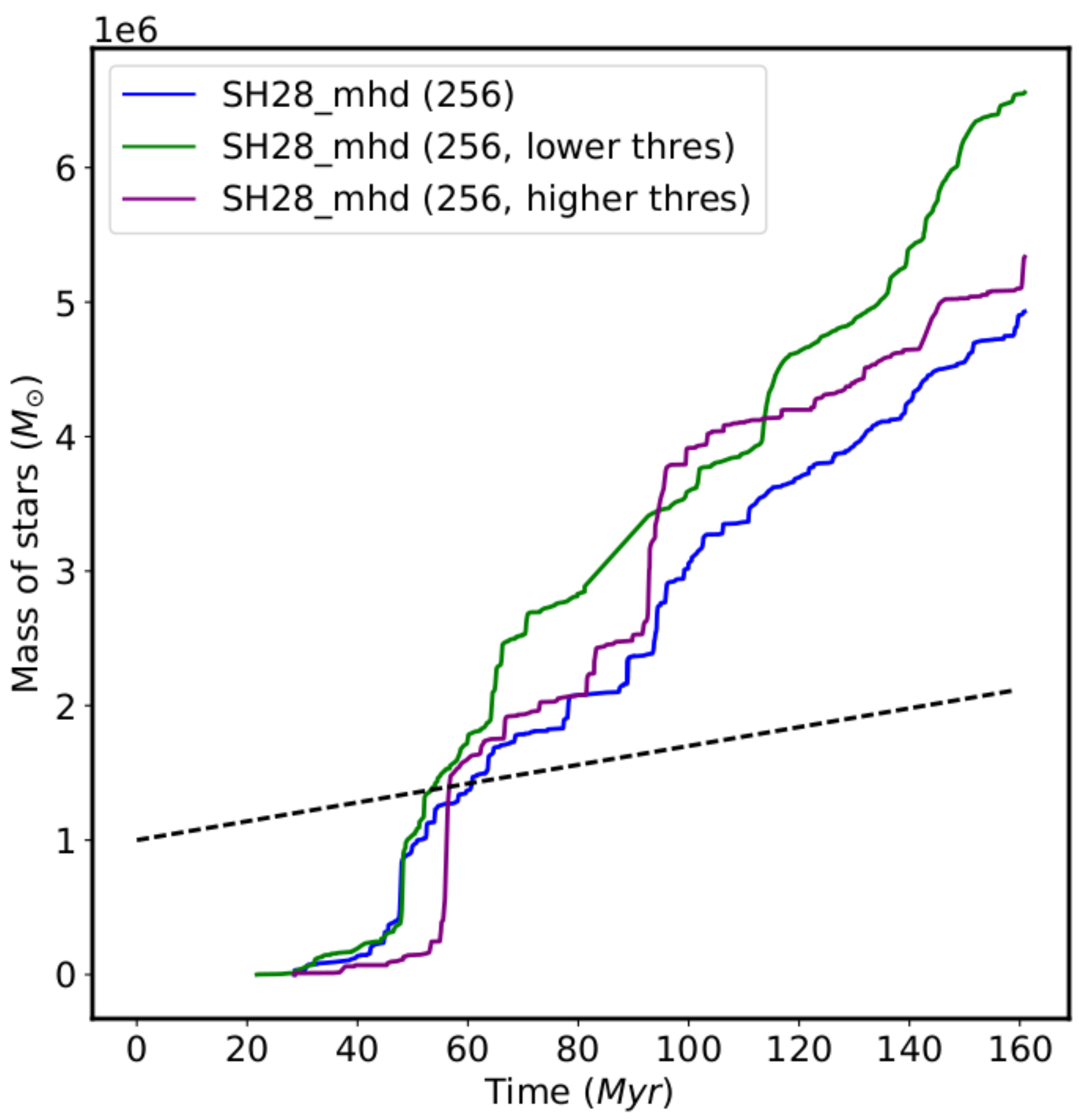}
    \caption{}
  \end{subfigure}
  \caption{Same as Fig. \ref{fig:plot_sfr1} but showing the influence of \textbf{numerical} parameters.
Top panel shows that the SFR does not really depend on the limitations on velocity and temperature of the gas 
and on the velocity of the massive stars in the sink particles.
Panel c displays the influence of the shear on simulations at higher resolution.
The last panels illustrate that star formation depends on the density threshold of the sink particles, giving a numerical incertitude factor of approximately two on the SFR.}
  \label{fig:plot_sfr2}
\end{figure*}

We now turn to Fig. \ref{fig:plot_sfr2} to study the influence of several numerical parameters on the star formation process. 

$\bullet$ \textbf{Panel (a)}:As explained in Sect. \ref{sec:sat}, simulations with higher values for $T_{sat}$ and $V_{sat}$ have been performed.
In these simulations, the total mass of stars formed is lower because the supernovae are more effective at injecting momentum into the ISM.
However, in the end, the asymptotic slopes (and then the asymptotic SFR) are quite similar.
Thus, at these values, these two parameters do not have a strong impact on the asymptotic SFR.
This justifies the choice of limiting the temperature and the velocity of the supernovae explosions that reduce the timestep.

$\bullet$ \textbf{Panel (b)}: As stated in Sect. \ref{sec:sat}, a simulation with $v_{disp} = 0$ has also been performed,
meaning that the massive stars leading to supernoave now stand motionless in the sink particle's frame.
In this simulation, the total mass of stars and the SFR is roughly the same as in the simulation with $v_{disp} = 1 \ \mathrm{km/s}$.
Thus, the value of this parameter has a negligible influence on the SFR. 
This is significantly different from the conclusion of \citet{Hennebelle14} (see also \citet{gatto2015}) and could stem
from the fact that support from galactic shear and HII regions is now considered, or that the supernovae can happen in 
more diffuse gas and be less effective because of the delay of the explosion.

\begin{figure}
  \begin{center}
    \includegraphics[width=0.47\textwidth]{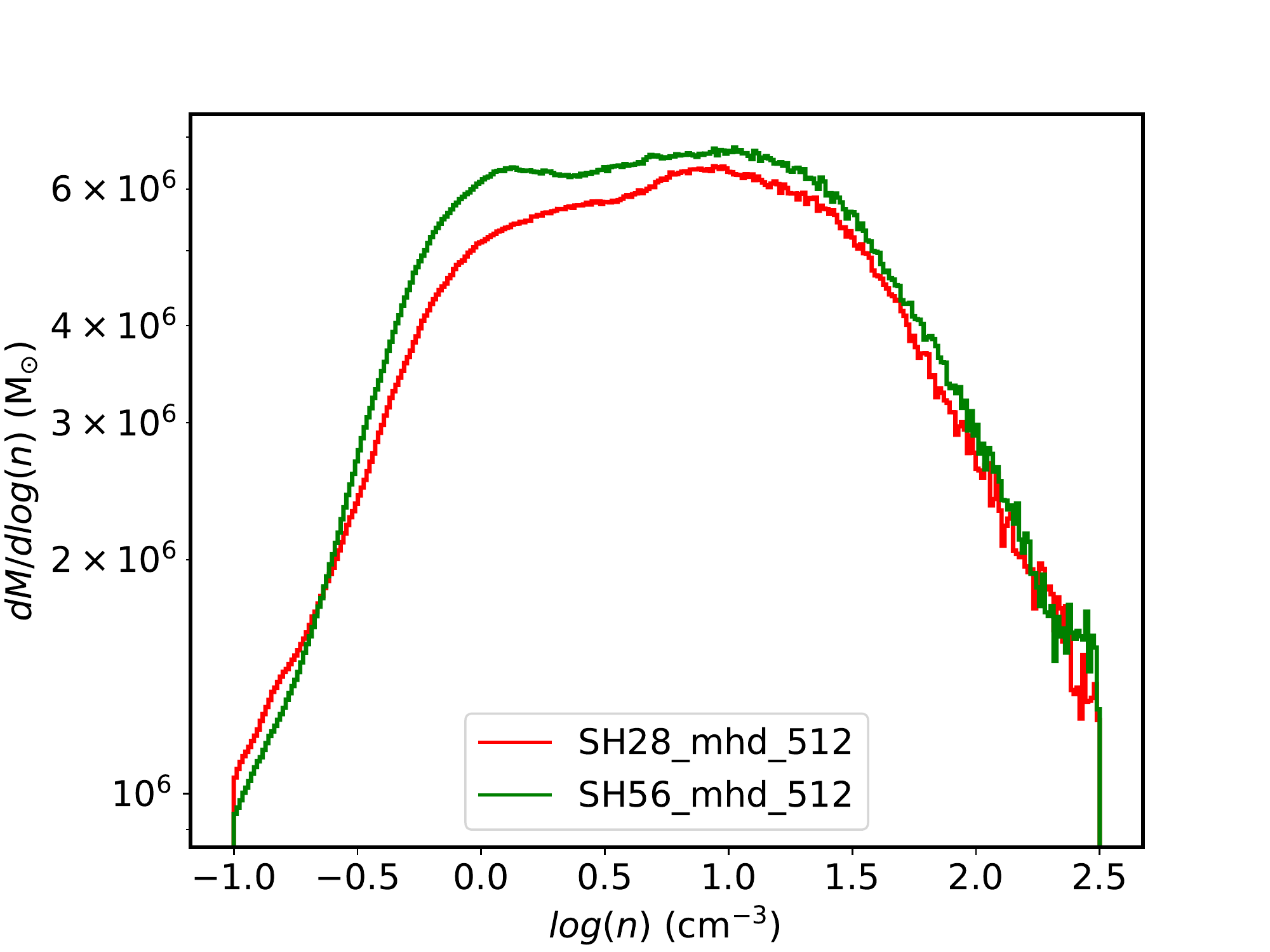}
  \end{center}
  \caption{Distribution functions of the density of the gas for MHD simulations (at $512^3$), at $80$ Myrs.
           \textbf{Red:} $V_{shear}=28 \ \mathrm{km.s^{-1}.kpc^{-1}}$. \textbf{Green:} $V_{shear}=56 \ \mathrm{km.s^{-1}.kpc^{-1}}$.}
  \label{fig:pdf}
\end{figure}

$\bullet$ \textbf{Panel (c)} shows that at a higher resolution, the dependence of the SFR on the shear is still the same: 
when the velocity gradient increases, the SFR decreases.
This result is consistent with the distribution functions of the gas for both simulations, which are simultaneously displayed in Fig. \ref{fig:pdf}.
The function is higher for greater shear simply because, as previously stated, the gravity is less efficient and less gas is accreted into the sink particles.
However, the pattern is the same for both simulations. \\
Going back to the SFR plots, while at $256^3$ the final slope/SFR of the green curve (SH56) matches the observed one (see the end of Sect. \ref{sec:sfr}), 
it is no longer the case at $512^3$. The final SFR is too high (by a factor of about two) at higher resolution for the column density used in this paper.
In order to explain this, we have performed complementary runs.

$\bullet$ \textbf{Panel (d)} 
shows the mass of stars formed for two identical simulations ($V_{shear} = 28 \ \mathrm{km.s^{-1}.kpc^{-1}}$),
except for the resolution ($256^3$ vs $512^3$, blue and red): the higher the resolution, the higher the mass, and the higher the SFR.
The problem is that the asymptotic behaviours do not match, the final SFR depends on the resolution 
(as stated above for the case $V_{shear} = 56 \ \mathrm{km.s^{-1}.kpc^{-1}}$): therefore it seems numerical convergence for the SFR has not been attained.  

However, it turns out that the numerical creation and accretion threshold of the sink particles (see Sect. \ref{sec:sink}) can explain why we do not have
numerical convergence of the SFR. Indeed, this parameter has been chosen until now to be constant and independent of the resolution ($10^3 \ \mathrm{cm^{-3}}$).
This threshold value was not well defined. \citet{Ostriker2017} argue that it should vary as
\begin{equation}
n_{sink} \propto \frac{1}{\Delta x^2}
\label{eq:thres}
,\end{equation}
where $\Delta x$ is the size of one cell. The quantity $\Delta x^2$ is then proportional to the surface of the cell.
Figure (d) displays a new simulation ($256^3$, green, with $n_{sink}$ four times lower (i.e. $250 \ \mathrm{cm^{-3}}$) 
than the threshold of the high resolution simulation (red). This way, the previous proportionality rule is respected.
With this prescription, the SFR of the low resolution case is now very close to that of the high resolution case, and thus numerical convergence is reached.

$\bullet$ \textbf{Panel (e)} shows the total mass of stars for the same resolution simulations but with three different sink thresholds:
$250 \ \mathrm{cm^{-3}}$, $10^3 \ \mathrm{cm^{-3}}$ , and $4.10^3 \ \mathrm{cm^{-3}}$.
The one with the higher $n_{sink}$ (purple) has some bursts and more mass, 
but the asymptotic SFR (from $100 \ \mathrm{Myrs}$) is actually lower, as one would expect.
It is, however, still higher than the observed one.
Between the two extreme cases, there is a factor of two in the asymptotic SFR.
The lower threshold case has an asymptotic SFR too high by a factor approximately six whereas the higher threshold case has an asymptotic SFR too high
by a factor of approximately three compared to the observed one.
This leaves an open question: even if numerical convergence is reached, 
what proportionality coefficient (or what reference value) must be taken in the numerical formula (\ref{eq:thres})? 
Without answering that question, there is still an incertitude of a factor  of two in the SFR.
This uncertainty should be understood as a limit to the present models.

\subsection{Star formation in simulations without virial criteria for the sink particles}
\label{sec:flags}

\begin{figure*}
  \begin{subfigure}[]{\columnwidth}
    \centering
    \includegraphics[width=7.1cm]{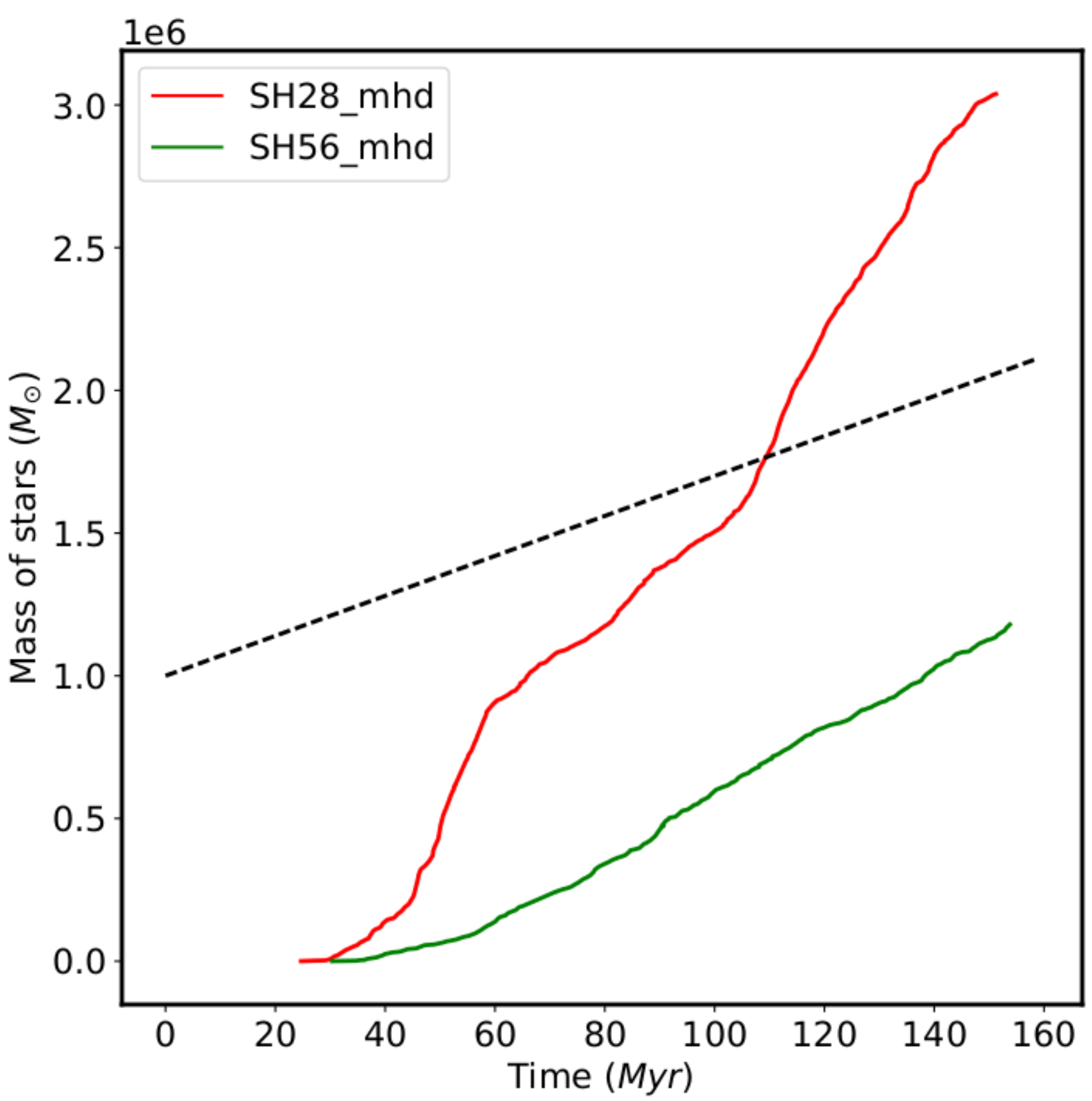}
    \caption{}
  \end{subfigure}
  \begin{subfigure}[]{\columnwidth}
    \centering
    \includegraphics[width=7.1cm]{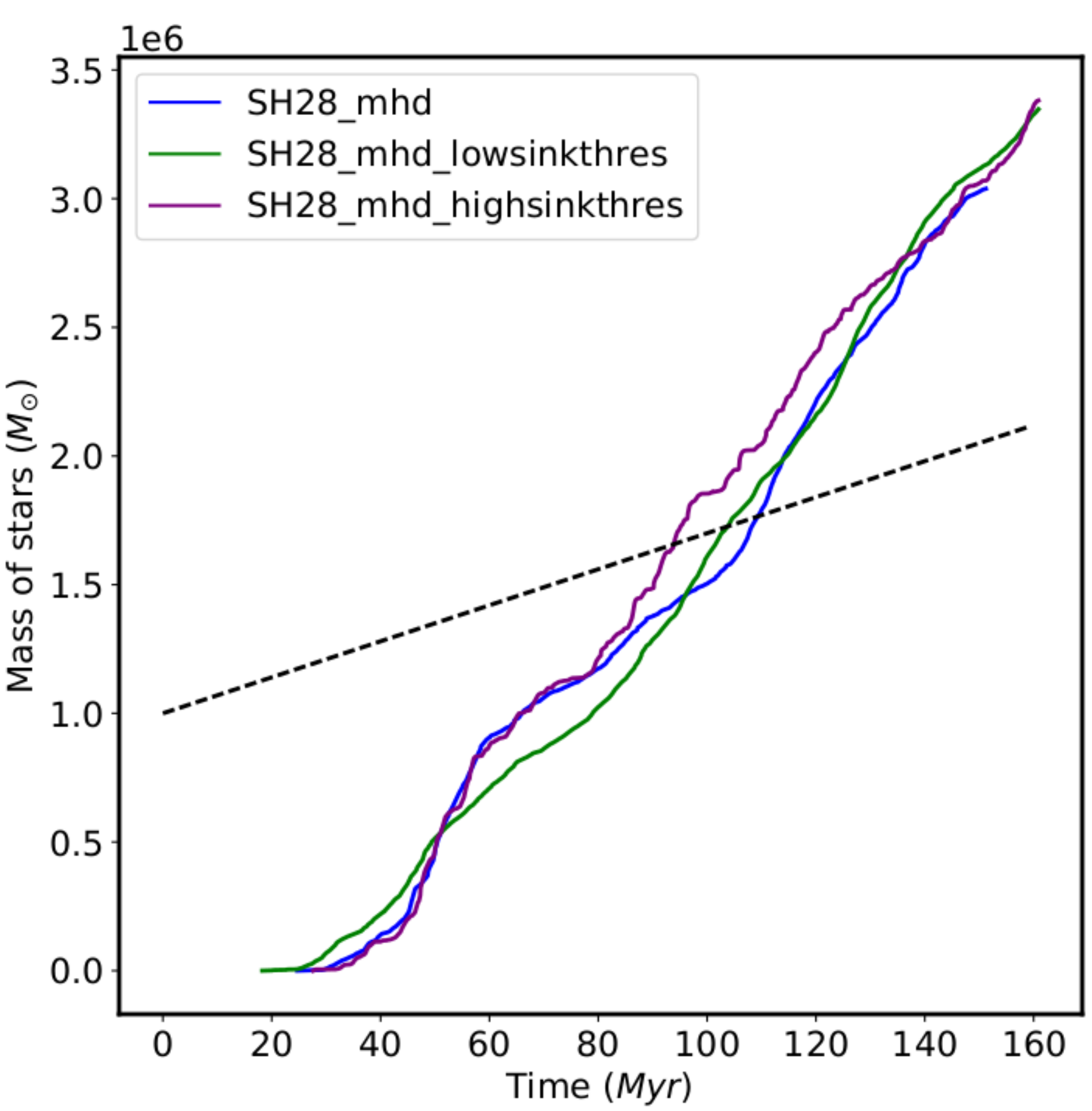}
    \caption{}
  \end{subfigure}
  \caption{Same as Fig. \ref{fig:plot_sfr1} but for simulations without the virial criteria for the sink particles.
Left panel shows that the total masses of stars are now lower but that the SFRs are similar to the ones measured in the runs with the virial tests.
Right panel shows that without the virial criteria, star formation no longer depends on the numerical density threshold of the sink particles.}
  \label{fig:plot_sfr3}
\end{figure*}

As explained in Sect. \ref{sec:sink}, we have performed runs without the virial criteria for the sink particles.
Figure \ref{fig:plot_sfr3} shows their star formation.

$\bullet$ \textbf{Panel (a)} shows the total mass of stars for two MHD runs, with both kinds of feedback, with
$V_{shear}=28$ or $56 \ \mathrm{km.s^{-1}.kpc^{-1}}$. Compared to the cases where the virial criteria are being tested
((e) of Fig. \ref{fig:plot_sfr1}), the total masses of stars are here lower by a factor of approximately two. 
This is because without the virial tests, the sink particles are created sooner, and thus stellar feedback takes effect sooner.
With the virial criteria, more mass of gas is accreted under gravity before the particles are formed, causing the many
bursts we see in Fig. \ref{fig:plot_sfr1} but not here in Fig. \ref{fig:plot_sfr3}.\\
The total masses are here lower, however the asymptotic SFRs are quite similar to the ones in the runs with the virial criteria.
When $V_{shear}=28 \ \mathrm{km.s^{-1}.kpc^{-1}}$ (red), the SFR is too high by a factor of approximately four compared to the observed one,
and when $V_{shear}=56 \ \mathrm{km.s^{-1}.kpc^{-1}}$, it is too high by a factor of $\sim 1.5$ as stated in Sect. \ref{sec:sfr}.

$\bullet$ \textbf{Panel (b)} shows the dependence of star formation on the threshold of the sink particles threshold $n_{sink}$.
The three curves are associated with the three previous values of $n_{sink}$: 
$250 \ \mathrm{cm^{-3}}$, $10^3 \ \mathrm{cm^{-3}}$ , and $4.10^3 \ \mathrm{cm^{-3}}$.
Contrary to the cases where the virial criteria must be satisfied ((e) of Fig. \ref{fig:plot_sfr2}), the total mass of stars and the SFR
do not depend significantly on the value of this numerical parameter.
This is likely because when virial criteria are considered, enough gas must have piled up to satisfy them. Here 
the sink particles are being introduced almost immediately and this makes the feedback less dependent on the 
system history.

\subsection{Structure properties}
\label{sec:structures}
\begin{figure*}
  \centering
  \includegraphics[width=0.49\textwidth]{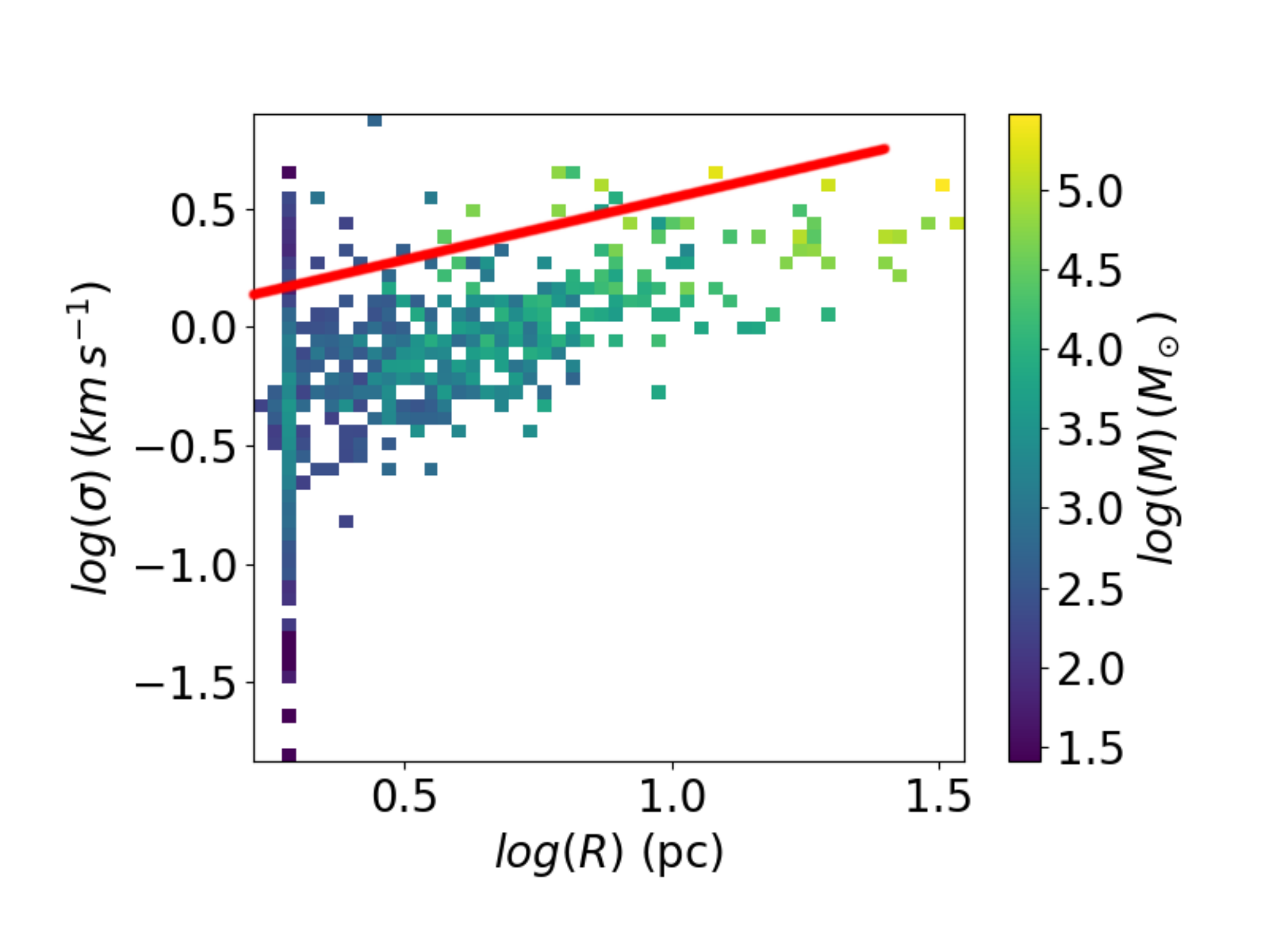}
  \includegraphics[width=0.49\textwidth]{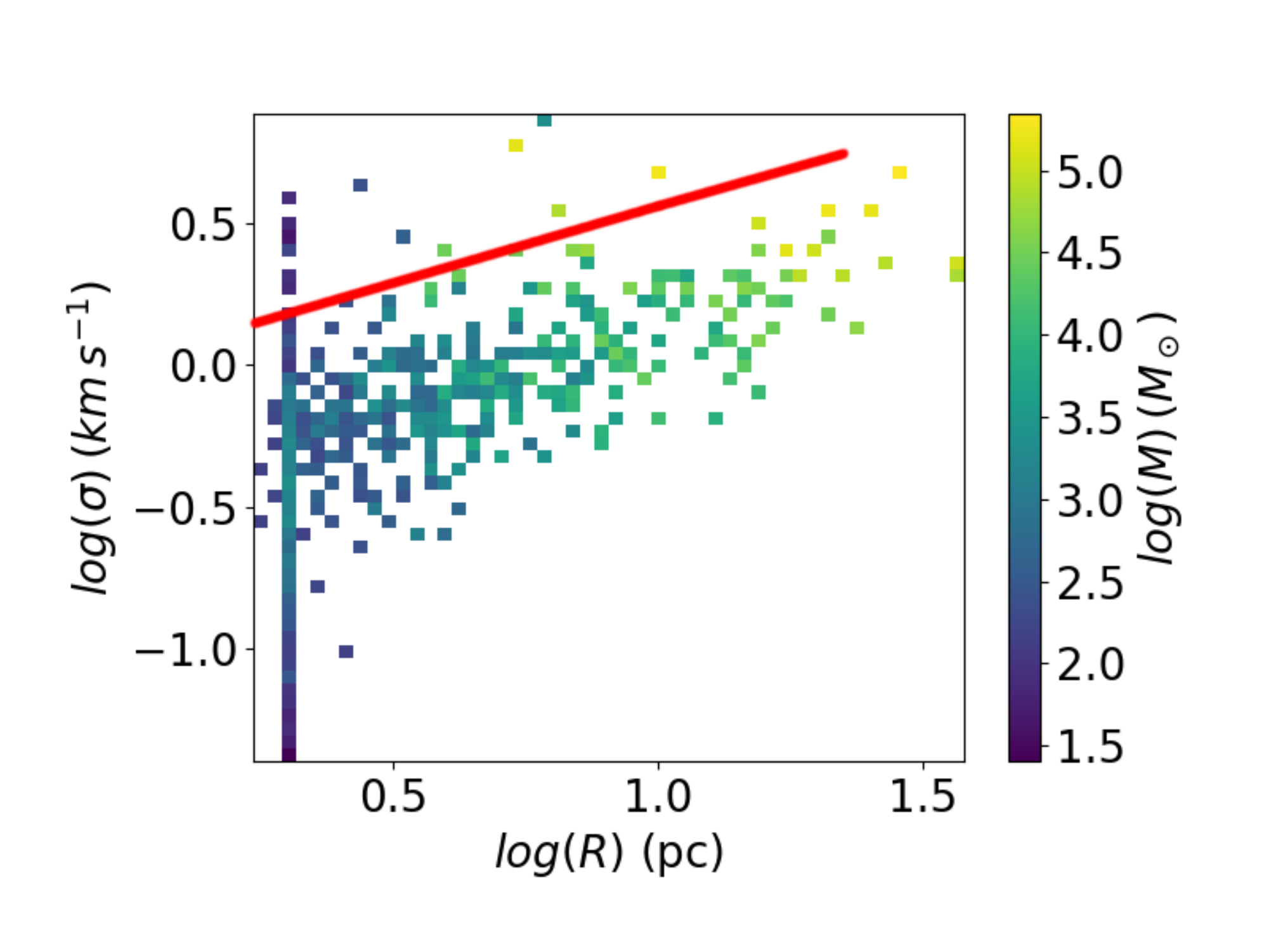}
  \caption{\textbf{Left:} Velocity dispersion - size relation of the clumps, at $80 \ \mathrm{Myrs}$ for the simulation SH28\_mhd\_512.
          \textbf{Right:} For the simulation SH56\_mhd\_512. Each square is a two-dimensional bin, and the colour depends on the total mass of the clumps associated with this bin.
           The red line is the observed Larson relation (Eq. \ref{eq:larsondis}), from \citet{Falgarone09}.}
  \label{fig:larsondis}
\end{figure*}

We now look at the properties of the dense clouds. The Larson relations \citep{Larson81} constitute observables that are worth reproducing.
More specifically, the relation between the velocity dispersion $\sigma$ in the clumps and their size $R$ is expected to be
\begin{equation}
\sigma = \sigma_0\left(\frac{R}{1 \ \mathrm{pc}}\right)^{\alpha}
\label{eq:larsondis}
.\end{equation}
From the data and the results of \citet{Falgarone09}, typical values of the parameters are
$\sigma_0=1.1 \ \mathrm{km/s}$ and $\alpha=0.5$.
These values are not well established in the literature (e.g. \citet{Heyer2009}); \citet{miville2016} infer a relation that depends
upon the column density of the cloud: $\sigma = 0.23 \; \mathrm{km/s} \; (\Sigma_HR)^{0.43}$).
However, a slightly different $\sigma_0$ or power index $\alpha$ would not change the following results.

In order to identify structures in our simulations, we proceed as in \citet{Iffrig2017} and 
use a friends to friends algorithm with a density threshold of $50 \ \mathrm{cm^{-3}}$. For each structure, the velocity dispersion $\sigma$ and the size $R$ are then computed like this:
\begin{align}
  \begin{split}
&M = \sum \rho dV \\
&\vec{v_0} = \frac{\sum \vec{v} \, \rho dV}{\sum \rho dV} \\ 
&\sigma^2 = \frac{1}{3} \frac{\sum (\vec{v}-\vec{v_0})^2 \, \rho dV}{\sum \rho dV} \\ 
&R = (\lambda_1 \lambda_2 \lambda_3 M^{-3})^{1/6},
  \end{split}
\end{align}
where the summations are done on the cells that define the clump.
The $\lambda_i$ are the eigenvalues of the inertia matrix defined in the centre of mass of the clump,
\begin{align}
  \begin{split}
&I_{11}=\sum (y^2+z^2) \, \rho dV, \ I_{22}=\sum (x^2+z^2) \, \rho dV, \\
&I_{33}=\sum (x^2+y^2) \, \rho dV, \ I_{12}=I_{21}= -\sum xy \, \rho dV, \\
&I_{13}=I_{31}= -\sum xz \, \rho dV, \ I_{23}=I_{32}= -\sum yz \, \rho dV. 
  \end{split}
\end{align}

Figure \ref{fig:larsondis} displays the velocity dispersions of the clumps versus their sizes,
for a velocity gradient $V_{shear}=28 \ \mathrm{or} \ 56 \ \mathrm{km.s^{-1}.kpc^{-1}}$.
The red straight line is the observed Larson relation (Eq. \ref{eq:larsondis}). The structures of our simulations follow a similar
power law, with a slope of the same order ($\alpha=0.5$). However, the velocity dispersions are well below the observed ones,
by a factor of approximately three, as already noted in \citet{Iffrig2017}.
This shows that the shear does not modify this conclusion and this may indicate that other energy sources are present and drive the turbulence,
like for example the large scale gravitational instabilities that are discussed in \citet{Krumholz15} and \citet{Krumholz17}.
They show that for massive galaxies with high gas column density and for rapidly-star-forming galaxies, 
gravity-driven turbulence is dominant over feedback-driven turbulence.
However, such large scale gravitational instability is absent in the model of this paper.

Another possible reason for these low velocity dispersions is that for smaller structures (with a size of a few cells), 
the dispersions are damped by numerical dissipation. However, even the structures larger than approximately ten cells
are below the expected velocity dispersion. Moreover, \citet{Iffrig2017} performed runs at various resolutions
and did not find significantly different dispersions between their more or less resolved runs
(see their Fig. 11 with runs B1 and B2L). If the low velocity dispersions were caused by numerical diffusion,
the high resolution simulations would have less dissipation for a given physical size and the velocity dispersions would be higher, but this is not the case.
Therefore, it seems more likely that this is actually a consequence of a missing injection of energy.

For small sizes ($\log(R) \lesssim 0.8$), the simulation with the highest shear (right)
has less clumps and they are less massive than in the simulation with the lowest shear (left).
For large sizes ($\log(R) \gtrsim 0.8$), the simulation with the highest shear (right) has more clumps and they are more massive than that of the lowest shear. This confirms what is stated in Sect. \ref{sec:shbox}: the galactic shear stretches the clouds 
(hence more large-sized structures and less small-sized structures at higher shear) and competes with gravity (hence less gas accreted into the stars).
\section{Discussion}
\label{sec:dis}

\subsection{Shear and column density}
\label{sec:sh_vs_density}

We have seen in Sects. \ref{sec:sfr} and \ref{sec:num} that the galactic shear competes with gravity and greatly reduces the SFR.
However, at the solar value $V_{shear}=28 \ \mathrm{km.s^{-1}.kpc^{-1}}$, the SFR was too high compared
to the observed Kennicutt value.
The reference value $V_{shear}$ in the solar neighbourhood may not be the correct choice to reproduce the Kennicutt relation
given our initial column density $\Sigma_{gas} = 19.1 \ \mathrm{M_{\odot} \, pc^{-2}}$.

\begin{figure}
  \centering
  \includegraphics[width=0.49\textwidth]{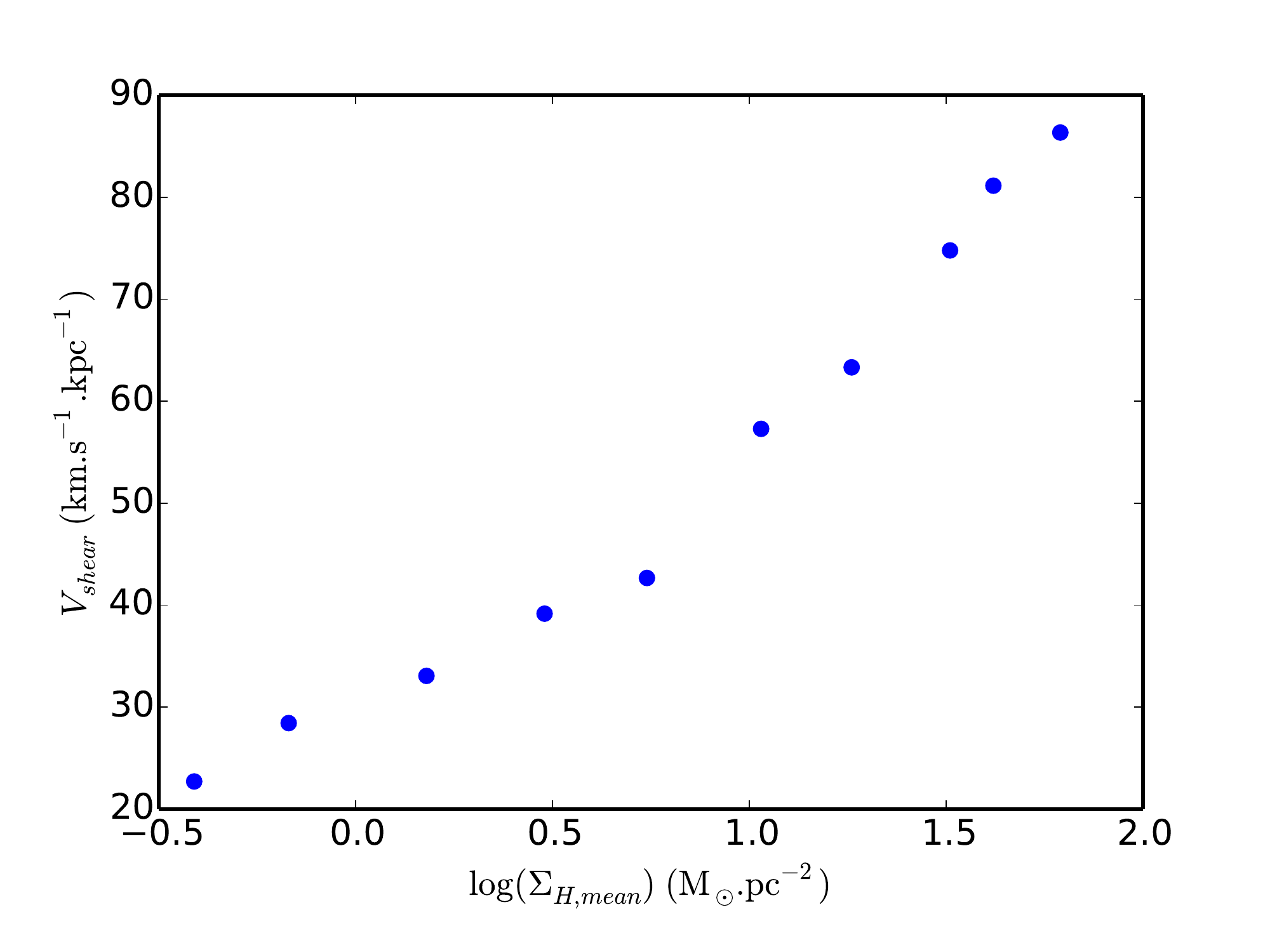}
  \caption{Scatter plot of Table \ref{tab:fred}.
          The mean value of the shear in the galaxy simulation of \citet{Bournaud2013} is given as a function of $\Sigma_{gas,mean}$.}
  \label{fig:sh_vs_density}
\end{figure}

Star-forming clouds are generally in spiral arms, where the shear is usually high \citep{Roberts69}.
In order to quantify this effect, we use simulations of a galaxy that is similar to the Milky Way, with the same radial density profile of the gas,
the same stellar mass distribution, and the same rotation curve. These simulations are presented in \citet{Bournaud2013} and \citet{Kraljic2014}. 
Figure 13 of \citet{Bournaud2013} shows the high shear in spiral arms: there can be a velocity gradient of $40 \ \mathrm{km/s}$ over $200 \ \mathrm{pc}$.
From these simulations, $1 \ \mathrm{kpc}$ boxes are isolated in the outer disk, between $5$ and $10 \ \mathrm{kpc}$ from the galactic centre.
This corresponds to regions located within or between spiral arms. 
From each one of these boxes, we compute the mass-weighted average of $\Sigma_{gas}$,
the usual surface-weighted average $\Sigma_{gas,mean}$ (which respects mass conservation), and the mass-weighted mean velocity gradient $V_{shear}$.
The surface-weighted $\Sigma_{gas}$ are used to define some bins, and the final results are presented in Table (\ref{tab:fred}).
Figure \ref{fig:sh_vs_density} is the corresponding scatter plot $V_{shear}$ versus $\log(\Sigma_{gas,mean})$.\\
The higher the column density, the higher is the velocity gradient. Then, for the column density of this paper, $\Sigma_{gas,mean} = 19.1 \ \mathrm{M_{\odot} \, pc^{-2}}$,
the relevant value should not be $V_{shear}=28 \ \mathrm{km.s^{-1}.kpc^{-1}}$ but about $V_{shear} \sim 63 \ \mathrm{km.s^{-1}.kpc^{-1}}$.
As previously stated (\emph{(d)} of Fig. \ref{fig:plot_sfr1} and \emph{(c)} of Fig. \ref{fig:plot_sfr2}), the SFR is very close to the observed one
for $V_{shear}=56 \ \mathrm{km.s^{-1}.kpc^{-1}}$. That shows that density and shear must not be taken independently, and that with a relevant value of $V_{shear}$
we are able to reproduce Kennicutt's law (within our intrinsic uncertainty of a factor approximately two).

We will explore the full parameter space of $\Sigma_{gas,mean}$ and $V_{shear}$ in a future work. 

\begin{table}
  \begin{center}
    \begin{tabular}{ccc}
\hline \hline
$\log{\Sigma_{gas}} \ (\mathrm{M_{\odot}.pc^{-2}})$ &$\log{\Sigma_{gas,mean}}$ & Mean $V_{shear} \ (\mathrm{km.s^{-1}.kpc^{-1}})$ \\
\hline
$-0.67 \cdots -0.33$ & $-0.41$ & $22.72$ \\
$-0.33 \cdots 0.00$ & $-0.17$ & $28.44$ \\
$0.00 \cdots 0.33$ & $0.18$ & $33.07$ \\
$0.33 \cdots 0.67$ & $0.48$ & $39.17$ \\
$0.67 \cdots 1.00$ & $0.74$ & $42.67$ \\
$1.00 \cdots 1.33$ & $1.03$ & $57.29$ \\
$1.33 \cdots 1.67$ & $1.26$ & $63.32$ \\
$1.67 \cdots 2.00$ & $1.51$ & $74.78$ \\
$2.00 \cdots 2.33$ & $1.62$ & $81.13$ \\
$2.33 \cdots 2.67$ & $1.79$ & $86.34$ \\
\hline
    \end{tabular}
  \end{center}
\caption{Mean gradient of velocities for different bins of $\log{\Sigma_{gas}}$. The values of $\Sigma_{gas}$ used to define these bins are mass-weighted averages
         whereas $\Sigma_{gas,mean}$ is the usual surface-weighted average. The mean $V_{shear}$ is mass-weighted.
         As can be seen, higher $V_{shear}$ are bound towards higher $\Sigma_{gas}$.
         Section \ref{sec:dis} gives more details.}
\label{tab:fred}
\end{table}

\subsection{Physical limits of the model}

The star formation efficiency (SFE) could be another key to restrain star formation: 
all the gas accreted onto the clusters is not necessarily converted into stars. 
A large fraction can be re-injected into the interstellar medium through stellar feedback operating at the cluster scale.
The SFE is defined for an embedded cluster as $M_{gas}/(M_{gas}+M_{stars})$ and its value is usually inferred to be around $ 20 \% $ \citep{LL03}.
However, as shown in \citet{lee2016a} this value is not well defined and should be regarded with great care.
In the simulations of this paper, the SFE is simply equal to $100 \%$. All the gas accreted onto the sink particles is considered as stars.
Having a model with a lower efficiency would obviously reduce the SFR.

The model of this paper also does not take stellar winds into account. Unlike supernovae, winds are not delayed,
and \citet{Gatto2017} find that they regulate the growth of young star clusters by suppressing the gas accretion on them 
shortly after the first massive star is born. 
In their model, when winds and supernovae are present, the SFR is close to the one with stellar winds only
because the gas surrounding the cluster is already unbound by the winds and supernovae have little additional effect.
They also find that adding stellar wind feedback to supernovae decreases $\Sigma_{SFR}$ by a factor of approximately two, compared to the case with supernovae feedback only.
Thus, stellar winds could be another missing physical process that would reduce the SFR even more in our model.

\section{Conclusions}

We have performed a set of simulations of $1 \ \mathrm{kpc^3}$ galactic regions, with differential rotation, magnetic field, and stellar feedback.
We have confirmed that the feedback from the stars has a huge limiting role on the star formation.
Our simulations also point out that with both HII regions and supernovae, there is a more
effective suppression of star formation than the linear combination of both effects would suggest.

As expected, the value of the galactic shear has a great impact: the higher the velocity gradient, the lower the SFR.
Using a galactic simulation, it seems this value actually depends on the column density of gas $\Sigma_{gas}$.
With a gradient $V_{shear}=56 \ \mathrm{km.s^{-1}.kpc^{-1}}$ and a column density $\Sigma_{gas}=19.1 \ \mathrm{M_{\odot} \, pc^{-2}}$, 
we get an asymptotic SFR very close to that of the observed Kennicutt law.

We manage to get numerical convergence of the SFR by making the threshold parameter of the sinks dependent on the spatial resolution:
$n_{sink} \propto 1/\Delta x^2$.
However, the proportionality coefficient is not well defined, and gives an incertitude of a factor of approximately two on the asymptotic SFR.

Finally, the analysis of the  structures of the properties confirmed that the galactic shear stretches the gas clouds.
We also find that the velocity dispersion is below the observed Larson law, by a factor  of approximately three.
This indicates that we may have missed some energy injection from the large galactic scales that drives the turbulence. We would expect such
an injection to reduce the SFR even more.

\begin{acknowledgements}
This work was granted access to HPC resources of CINES and CCRT  under the 
allocation  A0010407023 made by GENCI (Grand Equipement National de Calcul Intensif).
This research has received funding from the European Research Council under the European
Community's Seventh Framework Programme (FP7/2007-2013 Grant Agreement no. 306483).
We thank the referee for his helpful report.
\end{acknowledgements}

\bibliography{refs}{}
\bibliographystyle{aa}

\begin{appendix}
\section{Relevance of our gravitational boundary conditions}
\label{ap:bc}

\begin{figure}
  \centering
  \includegraphics[width=7.1cm]{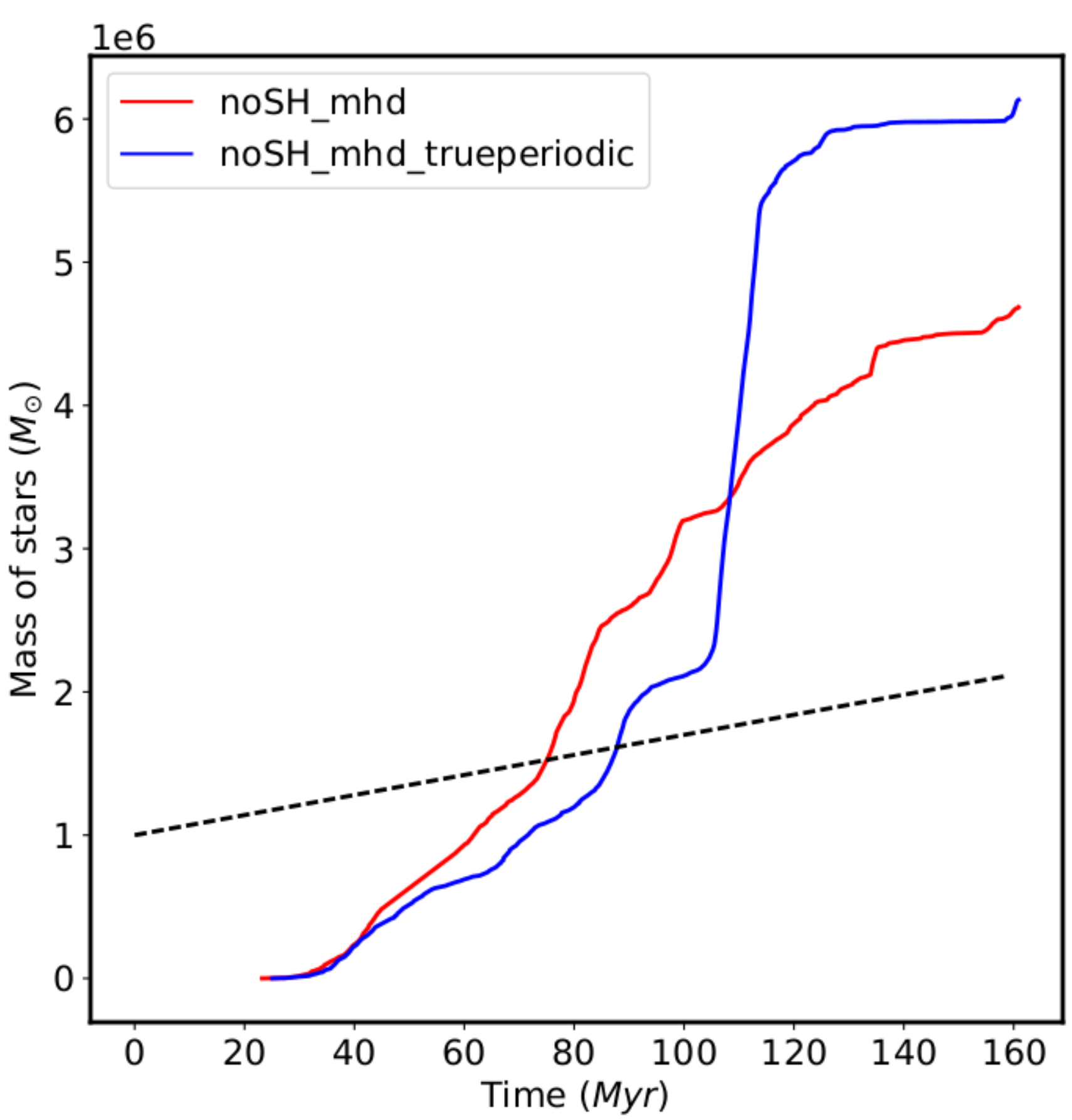}
  \caption{Star formation in the simulation ($V_{shear}=0$) with "shearing box" boundary conditions for gravity (Eq. \ref{eq:phi_bound})
compared with the same simulation with true periodic boundary conditions.}
  \label{fig:noflags_nosh}
\end{figure}

The boundary conditions used for gravity in Sect. \ref{sec:bcshbox} are still an approximation. 
For instance, as the boundary potential only depends upon $z$,
there is no force in the $x$ and in the $y$ directions. 
This can be troublesome, as gas and clouds are supposed to cross the boundary.
To test the influence of this approximation for the boundary conditions, we have performed a comparison 
between a fully periodic box and one with these conditions, for a run without shear ($V_{shear}=0$).
These runs were performed without the virial criteria of the sink particles (see Sect. \ref{sec:sink}).
The corresponding star formation curves are displayed in Fig. \ref{fig:noflags_nosh}.

The simulation of the fully periodic box (blue) has a jump of star formation at around $110 \ \mathrm{Myrs}$,
making it difficult to define a proper SFR.
However, the total mass of stars is about only $30 \%$ higher in the true periodic case at the end of the runs.
Then, estimating the slope from the end of the jump (at $115 \ \mathrm{Myrs}$), the case with the approximated
boundary conditions has a SFR higher by $\sim 40 \%$ compared to the fully periodic box. 
These differences 
 are lower than the numerical incertitude factor of two coming from the sink particles, so we neglected it in our discussion.
Moreover due to the large fluctuations that the SFR is experiencing, a finer quantification would require the realisation 
of many runs.  
One improvement of our numerical setup would be to define better boundary conditions for gravity when differential rotation is present.

\end{appendix}

\end{document}